\documentclass[a4paper,twocolumn,11pt,accepted=2021-05-20]{quantumarticle}
\pdfoutput=1
\usepackage[utf8]{inputenc}
\usepackage[english]{babel}
\pdfoutput=1
\usepackage[T1]{fontenc}
\usepackage{amsmath}
\usepackage{hyperref}
\usepackage{tikz}
\usepackage{lipsum}

\usepackage{geometry}
\usepackage{graphicx}  
\usepackage{bm}        
\usepackage{amssymb}   
\usepackage{xspace}
\usepackage{verbatim}
\usepackage{color}
\usepackage{soul}
\usepackage{subfigure}
\usepackage[export]{adjustbox}
\usepackage{dcolumn}
\usepackage{amsthm,stmaryrd,mathtools} 
\usepackage{txfonts}
\usepackage{braket}
\usepackage{amsmath}
\usepackage{xcolor}
\usepackage{array}
\usepackage{multirow}
\usepackage{tabularx}
\usepackage{booktabs}
\usepackage{lipsum}

\usepackage[numbers,sort&compress]{natbib}

\begin{document}

\title{Parallel entangling gate operations and two-way quantum communication in spin chains}

\author{Rozhin Yousefjani}
\orcid{0000-0002-1338-2031}
\email{RozhinYousefjani@uestc.edu.cn}
\author{Abolfazl Bayat}
\orcid{0000-0003-3852-4558}
\email{abolfazl.bayat@uestc.edu.cn}
\affiliation{Institute of Fundamental and Frontier Sciences, University of Electronic Science and Technology of China, Chengdu 610051, China}

\maketitle

\begin{abstract}
The power of a quantum circuit is determined through the number of two-qubit entangling gates that can be performed within the coherence time of the system. In the absence of parallel quantum gate operations, this would make the quantum simulators limited to shallow circuits. Here, we propose a protocol to parallelize the implementation of two-qubit entangling gates between multiple users  which are  spatially separated and use a commonly shared spin chain data-bus. Our protocol works through inducing effective interaction between each pair of qubits without disturbing the others, therefore, it increases the rate of gate operations without creating crosstalk. This is achieved by tuning the Hamiltonian parameters appropriately, described in the form of two different strategies. The tuning of the parameters, makes different bilocalized eigenstates responsible for the realization of the entangling gates between different pairs of distant qubits. Remarkably, the performance of our protocol is robust against increasing the length of the data-bus and the number of users.
Moreover, we show that this protocol can tolerate various types of disorders and is applicable in the context of superconductor-based systems.
The proposed protocol can serve for realizing two-way quantum communication.
\end{abstract}

\section{Introduction} 
\begin{figure}[t!]
\centering
\includegraphics[width=0.8\linewidth]{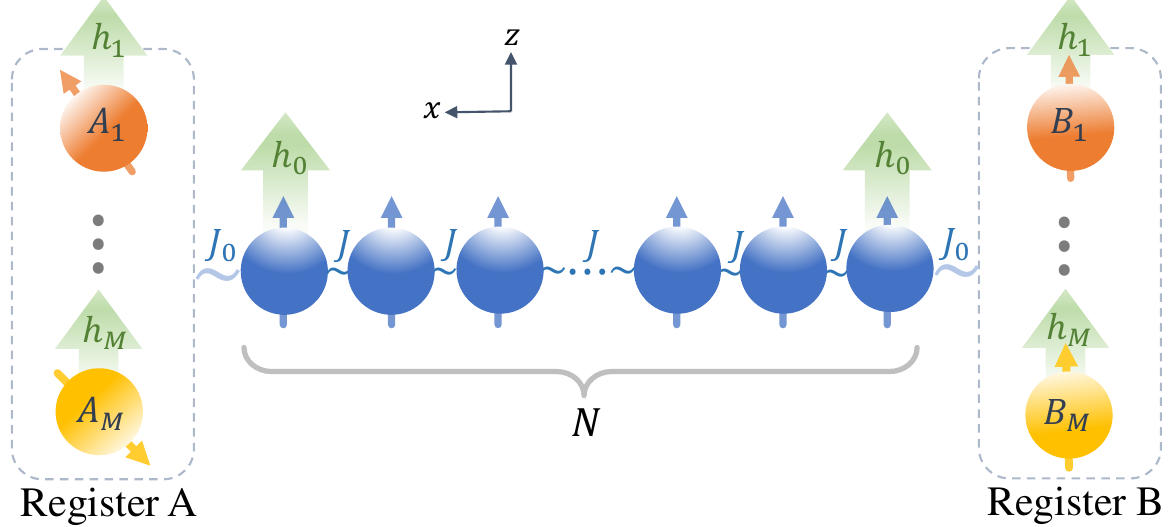}
\caption{ The schematic of simultaneous entangling gates between $M$-pair qubits of registers $A$ and $B$ across a spin chain as data-bus. By appropriately modulating the exchange coupling $J_0$, and local magnetic fields $h_0$ and $h_\nu$ ($\nu=1,\ldots,M$), each pair qubits $\{A_\nu,B_\nu\}$ would be mediated  with a different set of system's energy levels.}\label{fig:Schematic}
\end{figure} 
In order to achieve universal computation on a digital quantum simulator one requires the capability of performing arbitrary local single-qubit unitary rotations on every qubit as well as one type of two-qubit entangling gate between any pair of qubits~\cite{PhysRevA.52.3457}. The single-qubit unitary operations are performed locally through external control fields and have been implemented with very high fidelity in various physical setups. The two-qubit entangling gate, however, can only be realized through interaction between the two qubits~\cite{PhysRevLett.89.247902} and have been realized in quantum dots~\cite{shulman2012demonstration,Veldhorst2015}, dopant-based systems~\cite{He2019}, optical lattices~\cite{mandel2003controlled,bloch2008quantum}, ion traps~\cite{Schafer2018,PhysRevLett.117.060504,harty2014high,ballance2016high,gaebler2016high}, super conducting devices~\cite{Barends2014,foxen2020demonstrating}, Rydberg atoms~\cite{Yu2019} and diamond nitrogen-vacancy centers~\cite{PhysRevLett.122.010503}. The demand for direct interaction makes the realization of two-qubit gates very challenging for distant qubits. Thus, several proposals have been put forward to mediate the interaction between distant qubits using a shuttled particle~\cite{Bertrand2016,Fujita2017,Mills2019}, a traveling wave packet~\cite{Bienfait2019,moehring2007entanglement,togan2010quantum,PhysRevLett.120.200501}, a shared spatially extended mode~\cite{cirac1995quantum,rabl2010quantum} or a spin chain data-bus~\cite{bose2003quantum,bayat2010information,yang2011entanglement,yao2011robust}.
The latter, namely spin chain setups~\cite{Ronke2011,blundell2004organic,zwanenburg2013silicon,fazio2001quantum,porras2004effective,blatt2012quantum}, are particularly useful for mediating the interaction between two distant qubits as they are made from the same physical systems as the logical qubits and hence
eliminate the adversity of interfacing between different physical systems. The dynamics of spin chain systems have already been harnessed to implement different quantum gates between spatially separated qubits~\cite{banchi2011nonperturbative,yao2011robust,yung2005perfect,
yung2006processor,gorshkov2010photonic,weimer2012long,Estarellas2017}.

One of the main challenges in current quantum simulators is the finite coherence time which restricts the total number of gates that can operate. In addition, many implementations of the two-qubit gates allow for only one or very few gates at each instance. This substantially reduces the operation rate of quantum processors and restricts their ability to realize deep circuits. To overcome this obstacle, in the context of state transfer, several ideas have been developed to mediate interactions between multiple qubits~\cite{apollaro2015many,chetcuti2020perturbative} or exploit dense coding like ideas in spin systems~\cite{yang2011entanglement}.  
Current classical computers benefits form parallel computations by exploiting Multiple Instruction, Multiple Data (MIMD) architectures.
This boosts their computational power while increasing the frequency scaling of their processors is practically impossible. 
Likewise, a quantum version of MIMD is highly desirable to design new protocols that are able to implement multiple entangling gates in parallel and enhance the operation rate within the coherence time of the hardware. Quantum gate parallelism which is essential for fault-tolerant error correction~\cite{Aharonov2008,Steane1998}  
has so far been realized in ion-traps~\cite{Figgatt2019,Grzesiak2020}, superconducting circuits~\cite{Song2017} and optical lattices~\cite{mandel2003controlled,Levine2019}. Nonetheless, the development of parallel operation of two-qubit gates between \emph{selected} pair of qubits in the context of spin-based computation has remained a critical open question.   

In this paper, we address this problem and put forward a protocol that  implements parallel multiple two-qubit entangling gates on several distant pairs of qubits using a shared spin chain data-bus. The same setup can also be used for realizing two-way quantum communication which shows significant improvement over previous proposals~\cite{PhysRevA.84.022345}. The idea of this work is based on our previous work~\cite{yousefjani2019simultaneous} which accelerates the rate of communication in quantum networks by allowing multiple users to simultaneously communicate through a common spin chain channel. To achieve parallel gate operation, we  create an effective interaction between each pair of users through properly tuning the Hamiltonian parameters. This is achieved through two different strategies which optimize different set of Hamiltonian parameters.
Although our proposal is general and can be implemented in various
physical platforms, we exclusively propose an
application based on superconductor qubits. 
Remarkably, our protocol shows acceptable robustness against fabrication imperfections and quantum noise effects including dephasing and amplitude damping.  
   
\section{Model}
We consider an array of $N$ spin-$1/2$ particles as our data-bus in which particles interact via XX Hamiltonian  
\begin{equation}\label{eq:channel Hamiltonian}
H_{ch}=J \sum_{i=1}^{N-1}(\sigma_{i}^{x}\sigma_{i+1}^{x}+\sigma_{i}^{y}\sigma_{i+1}^{y})+h_0(\sigma_{1}^{z}+\sigma_{N}^{z}),
\end{equation}
where $\sigma_{i}^{x,y,z}$ are the Pauli operators acting on site $i$, $J$ is the spin exchange coupling and $h_0$ represents the transverse magnetic field acting only on the end sites. 
We assume this spin-chain is shared between two remote quantum registers $A$ and $B$, each containing $M$ spin qubits labeled by $A_\nu$ and $B_\nu$ ($\nu=1,\cdots,M$), see Fig.~\ref{fig:Schematic}.
The interaction between the registers' qubits and the data-bus is given by 
\begin{eqnarray}\label{eq:Interaction Hamiltonian}
H_I&=&J_0 \sum_{\nu=1}^{M}\left(\sigma_{A_\nu}^{x}\sigma_{1}^{x}+\sigma_{A_\nu}^{y}\sigma_{1}^{y}+\sigma_{N}^{x}\sigma_{B_\nu}^{x}+\sigma_{N}^{y}\sigma_{B_\nu}^{y} \right)\cr \cr
&+&
\sum_{\nu=1}^{M} h_{\nu}(\sigma_{A_\nu}^{z}+\sigma_{B_\nu}^{z}),
\end{eqnarray}
where $J_0$ denotes the coupling between the registers and the data-bus and $h_\nu$ is the transverse magnetic field applying on the pair spin qubits $\{A_\nu,B_\nu\}$.
We assume that qubits of the register $A$ ($B$) are initially prepared in the normalized states $\vert \psi_{\nu} \rangle_{A}=\alpha^{0}_{\nu}|0_{\nu}\rangle+\alpha^{1}_{\nu}|1_{\nu}\rangle$ 
($\vert \varphi_{\nu} \rangle_{B}=\beta^{0}_{\nu}|0_{\nu}\rangle+\beta^{1}_{\nu}|1_{\nu}\rangle$), and are decoupled from the data-bus which is initialized in the state $\vert \bm{0} \rangle_{ch}{=}|0,\ldots,0\rangle_{ch}$. 
Therefore, the state of the whole system becomes  
\begin{equation}\label{eq:Initial state}
\vert \Psi(0) \rangle = |\psi_{1},\ldots,\psi_{M}\rangle_{A} |\bm{0}\rangle_{ch} |\varphi_{1},\ldots,\varphi_{M}\rangle_{B}.
 \end{equation}
Once the coupling $J_{0}$ is switched on at time $t{=}0$, this quantum state evolves as $\vert \Psi(t)\rangle {=} e^{-iHt} \vert \Psi(0)\rangle$, where $H {=} H_{ch} {+} H_{I}$ is the total Hamiltonian of the system.
\begin{figure*}[t!]
	\centering\offinterlineskip
	\includegraphics[width=0.48\linewidth, height=4.2cm]{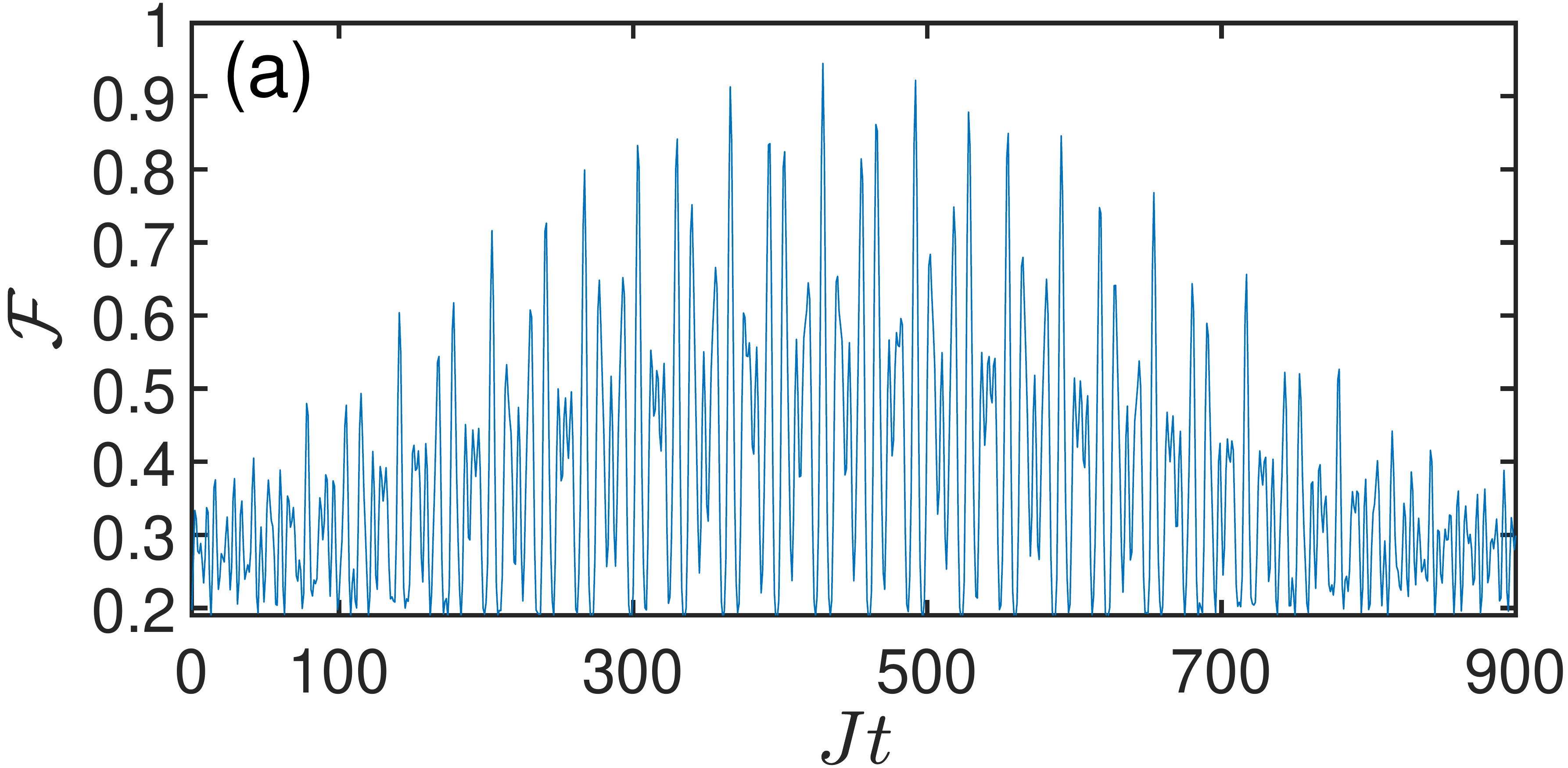}
	\includegraphics[width=0.48\linewidth, height=4.2cm]{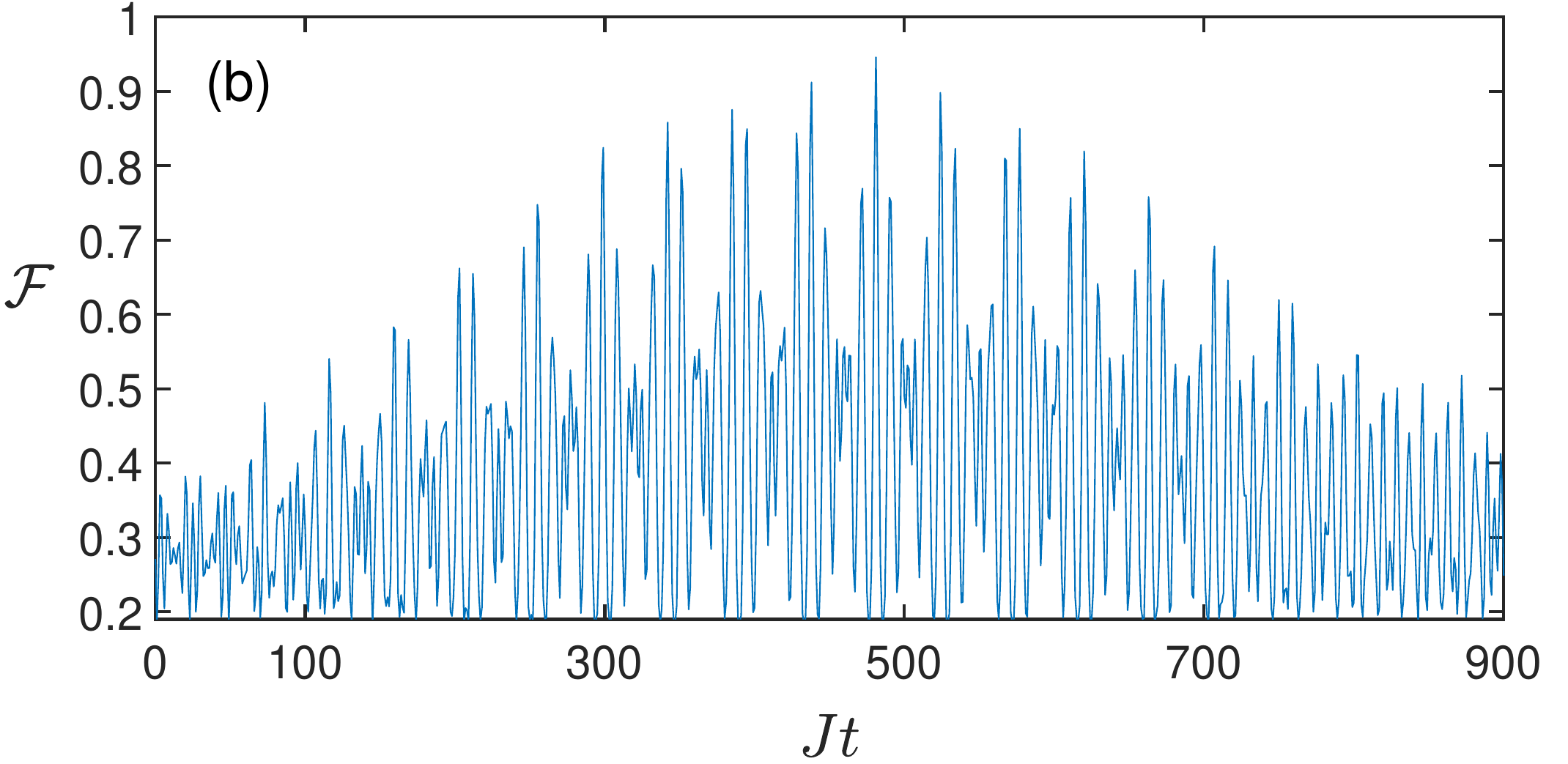}
	\caption{$M{=}2$: The average $\mathcal{F}{=}(\overline{F}_{1}+\overline{F}_{2})/2$ for our two strategies $\bm{S1}$ (a) and $\bm{S2}$ (b) as a function of time in a chain of $N{=}20$.  The Hamiltonian parameters for $\bm{S1}$ and $\bm{S2}$ are taken as $\{J^{opt}_{0}/J{=}0.04, h^{opt}_{1}/J{=}0.35, h^{opt}_{2}/J{=}{-}0.25\}$ and $\{ h^{opt}_{0}/J{=}25, h^{opt}_{1}/J{=}0.4, h^{opt}_{2}/J{=}{-}0.25\}$, respectively.}\label{fig:AGF vs time}
\end{figure*}
In Ref.~\cite{yousefjani2019simultaneous}, a protocol for simultaneous quantum communication between multiple users across a shared spin chain data-bus was proposed. In that protocol, one can achieve simultaneous high-fidelity state transfer between qubit pairs $\{A_{\nu},B_{\nu}\}$, with low crosstalk, through  appropriately tuning the local Hamiltonian parameters, namely $J_0$, $h_0$, and $h_\nu$. Such tuning  creates bilocalized eigenstates between each pair of users, namely qubits $\{A_{\nu},B_{\nu}\}$, which then mediate direct interaction between them  without affecting the others.
According to Ref.~\cite{yousefjani2019simultaneous}, the tuning of the parameters for simultaneous state transfer requires the following steps:
\begin{itemize}
	\item[(I)] Establishing an effective end-to-end interaction, i.e. confining the excitations to the  qubits of the registers and leaving the channel approximately unexcited, i.e. $|\bm{0}\rangle_{ch}$, at all times, by either decreasing $J_0$ \cite{bayat2015measurement,Apollaro2020,wojcik2005unmodulated,venuti2006long,venuti2007long,paganelli2013routing} or increasing $h_0$ ~\cite{lorenzo2013quantum,apollaro2015many} or both.
   \item[(II)] Making each pair of qubits $\{A_{\nu},B_{\nu}\}$ off-resonant from the others through tuning the local magnetic fields $h_{\nu}$.
\end{itemize}
Here, we extend these results to perform parallel multiple two-qubit entangling gates
on pairs of qubits in the registers $A$ and $B$. We also find out that the same protocol can be used for two-way communication.  

In the case of $M{=}1$, the condition (I) and the free fermionic nature of the model results in dynamics which at a certain time $t=\tau$ can be well approximated as~\cite{banchi2011nonperturbative,yang2011entanglement}
\begin{equation}\label{eq:evolution_1pair}
    e^{-iH\tau}|a_1\rangle_A|\bm{0}\rangle_{ch}|b_1\rangle_B \simeq e^{i\phi_{a_1 b_1}^1} |b_1\rangle_A|\bm{0}\rangle_{ch}|a_1\rangle_B.
\end{equation}
Remarkably, for different choices of $a_1,b_1=0,1$, at time $t=\tau$, the phases $\phi_{ab}$ take values such that~\cite{banchi2011nonperturbative}   
\begin{equation}\label{eq:phases}
   \phi_{00}^1{=}0, \hspace{4 mm} 
   \phi_{01}^1{=}\phi_{10}^1{=}(N+1)\pi/2, \hspace{4 mm}
   \phi_{11}^1{=}N\pi,
\end{equation}
where $\phi_{00}^1$ is taken to be zero as the reference and  $\phi_{01}^1{=}\phi_{10}^1$ is guaranteed due to the mirror symmetry of the system. Therefore, this dynamics performs a quantum gate $G_1$ between the two qubits of the registers $A$ and $B$ 
\begin{equation}\label{eq:gate_1pair}
    G_1|a_1\rangle_A |b_1\rangle_B \cong e^{i\phi_{a_1 b_1}^1} |b_1\rangle_A |a_1\rangle_B.
\end{equation}
This gate  not only swaps the qubits of the registers, but also imprints a phase which depends on the initial state of the qubits.
The resulted phases at time $t=\tau$, given in Eq.~(\ref{eq:phases}), makes $G_1$ an entangling gate which creates a maximally entangled state between the two qubits if they start with $|\psi_\nu\rangle_A|\phi_\nu\rangle_B=|+\rangle_A|+\rangle_B$, where $|+\rangle=(|0\rangle+|1\rangle)/\sqrt{2}$. The goal of this paper is to generalize these results to multiple users, namely $M{>}1$, where the dynamics performs several two-qubit gates  on the pairs  $\{A_\nu,B_\nu\}$ (for $\nu=1,\cdots,M$) in parallel. 
In the case of $M>1$, the initial state of Eq.~(\ref{eq:Initial state}) takes the form
\begin{equation}\label{eq:Initial state1}
\vert \Psi(0) \rangle = \sum_{\bm{a},\bm{b}}\alpha_{\bm{a}}\beta_{\bm{b}} \:|\bm{a}\rangle_{A}|\bm{0}\rangle_{ch}|\bm{b}\rangle_{B},
 \end{equation}
where vectors $|\bm{a}\rangle_{A}{=} |a_1,\ldots,a_M\rangle_{A}$ and $|\bm{b}\rangle{=} |b_1,\ldots,b_M\rangle_{B}$ with $a_{\nu},b_{\nu}{=}0,1$, denote the computational basis of the registers $A$ and $B$, respectively, and $\alpha_{\bm{a}}= \prod_{\nu =1}^{M} \alpha_{\nu}^{a_{\nu}}$ and $\beta_{\bm{b}}=\prod_{\nu =1}^{M} \beta_{\nu}^{b_{\nu}} $ are abbreviations for multiplied coefficients of the initial states.
Notably, satisfying the condition (I) leads to the emergence of bilocalized eigenstates whose excitations are mainly localized at the sites of the registers' qubits. 
These bilocalized eigenstates mediate the coupling between the computational states of the registers.  
By applying local magnetic field $h_\nu$ and meeting the condition (II), the excitations would be more localized between only two qubits, namely $A_\nu$ and $B_\nu$  (see Appendix C in Ref~\cite{yousefjani2019simultaneous}).
This can be achieved by properly optimizing  $h_{\nu}$'s to be adequately far from each other. 
Since the bilocalized eigenstates
are the only ones involving in the dynamics of the system, each qubit pair $\{A_{\nu},B_{\nu}\}$ evolves without disturbing the others and the channel mostly remains unexcited.
In that case, the dynamics of the system at special time $t=\tau$  leads to    
\begin{equation}\label{eq:time evolution of basis}
e^{-iH\tau}|\bm{a}\rangle_{A}|\bm{0}\rangle_{ch}|\bm{b}\rangle_{B}\simeq e^{i\Phi_{\bm{ab}}}|\bm{b}\rangle_{A}|\bm{0}\rangle_{ch}|\bm{a}\rangle_{B},
\end{equation}     
where $\Phi_{\bm{ab}}{=}\sum_{\nu=1}^M \phi_{a_\nu,b_\nu}^\nu$ and again the mirror symmetry implies $\Phi_{\bm{ab}}=\Phi_{\bm{ba}}$.
This state inversion allows us to introduce a global gate $\bm{G}$ between registers $A$ and $B$ as
\begin{equation}\label{eq:global_gate}
\bm{G}|\bm{a}\rangle_{A}|\bm{b}\rangle_{B}\simeq e^{i\Phi_{\bm{ab}}}|\bm{b}\rangle_{A}|\bm{a}\rangle_{B}.
\end{equation} 
The special form of $\Phi_{\bm{ab}}$ allows to write  $\bm{G}\simeq G_{1}G_{2}\ldots G_{M}$, where $G_{\nu}$, is the two-qubit gate which acts on pair $\nu$, similar to Eq.~(\ref{eq:gate_1pair}). 
The evolution in Eqs.~(\ref{eq:evolution_1pair}) and~(\ref{eq:time evolution of basis}) are very ideal and in reality, there are two main issues which deviate this perfect picture. The first one is the small dispersion in the system which leaks some information to the channel~\cite{lewis2019dynamics,eisert2015quantum}. The second issue is that the cross talk is not exactly zero and some information may leak to other pairs. These effects induce some entanglement between the data-bus and the registers, preventing the gate $\bm{G}$ and consequently $G_{\nu}$'s from being  perfect unitary operations.
In that case, the dynamics of each pair $\{A_{\nu},B_{\nu}\}$ should be considered as a completely positive and trace preserving map, $\rho_{\nu}(t){=}\Lambda^{\nu}(t)[\rho_{\nu}(0)]$. 
Here, $\Lambda^{\nu}(t)[\rho_{\nu}(0)]{=}Tr_{\widehat{\nu}}(\vert \Psi(t)\rangle\langle \Psi(t)\vert)$ in which $Tr_{\widehat{\nu}}$ means trace over all qubits except the pair  $\{A_\nu , B_\nu \}$. 
To measure how well the map $\Lambda^{\nu}(t)$ approximates each two-qubit gate $G_\nu$, one can use the average gate fidelity~\cite{nielsen2002simple}
\begin{equation}\label{Average Gate Fidelity1}
\overline{F}_{\nu}(t)=\int d\psi \langle \psi \vert G_\nu^{\dagger} \Lambda^{\nu}(t)[\vert \psi\rangle\langle \psi \vert] G_\nu\vert \psi\rangle, 
\end{equation}
where the integral is over the uniform (Haar) measure
$d\psi$ on two-qubit state space, normalized as $\int d\psi=1$. Rewriting Eq.~(\ref{Average Gate Fidelity1}) in the tow-qubit computational basis, combined with some straightforward calculations, leads to
\begin{equation}\label{Average Gate Fidelity2}
\overline{F}_{\nu}(t)=\dfrac{1}{5}+\dfrac{1}{20}\sum_{ii^{\prime}jj^{\prime}}(G_{\nu}^{*})_{ij}\langle i \vert \Lambda^{\nu}(t)[\vert j \rangle \langle j^{\prime} \vert] \vert i^{\prime} \rangle (G_{\nu})_{i^{\prime}j^{\prime}}.
\end{equation}
Our goal is to maximize the average gate fidelity $\overline{F}_{\nu}$ for all pairs $\{A_{\nu},B_{\nu}\}$ at the same time.
This can be pursue by maximizing the average  $\mathcal{F}{=}\sum_{\nu=1}^{M}\overline{F}_{\nu}/M$ via controlling the Hamiltonian parameters $J_{0}$, $h_{0}$ and $h_{\nu}$'s.
Our protocol can be established in two different strategies based on the set of the Hamiltonian parameters which are chosen to be optimized.
In our first strategy, labeled by $\bm{S1}$, we set $h_0{=}0$ and attempt to create effective end-to-end interaction via optimizing $J_0{<}J$.
In the second strategy, $\bm{S2}$, this effective interaction would be induced by applying strong magnetic field $h_0{>}J$ on the end sits of the data-bus while the coupling are kept uniform, i.e., $J_0{=}J$.
Each of these strategies might be suitable for a different physical platform. 
Throughout the paper and for both strategies, we fix a 
time window, for the dynamics of the system and then maximize the average gate fidelity $\mathcal{F}$ with respect to the Hamiltonian
parameters to find their optimal values, namely $J_{0}^{opt}/J$, $h_{0}^{opt}/J$ and $h_{\nu}^{opt}/J$, by brute-force optimization method. 
In fact, tuning the Hamiltonian parameters to these optimal values establishes an effective interactions between registers' qubits and, hence, the gate operation between each pair of qubits takes place with the highest quality at a special time $\tau$.
For the sake of clarity, the average gate fidelity $\mathcal{F}$ that is obtained for the optimal parameters and the desired gate duration $\tau$ is denoted as $\mathcal{F}^{max}{=}\sum_{\nu=1}^{M}\overline{F}^{max}_{\nu}(\tau)/M$.
In the following, we first restrict ourselves to the case of $M{=}2$, and evaluate the performance of two strategies $\bm{S1}$ and $\bm{S2}$. Then, we extend the results to larger $M$. 
\subsection{Parallel gate operation for $M{=}2$} 
\begin{figure*}[t!]
\begin{minipage}{2\columnwidth}
  \includegraphics[width=8cm,height=4cm]{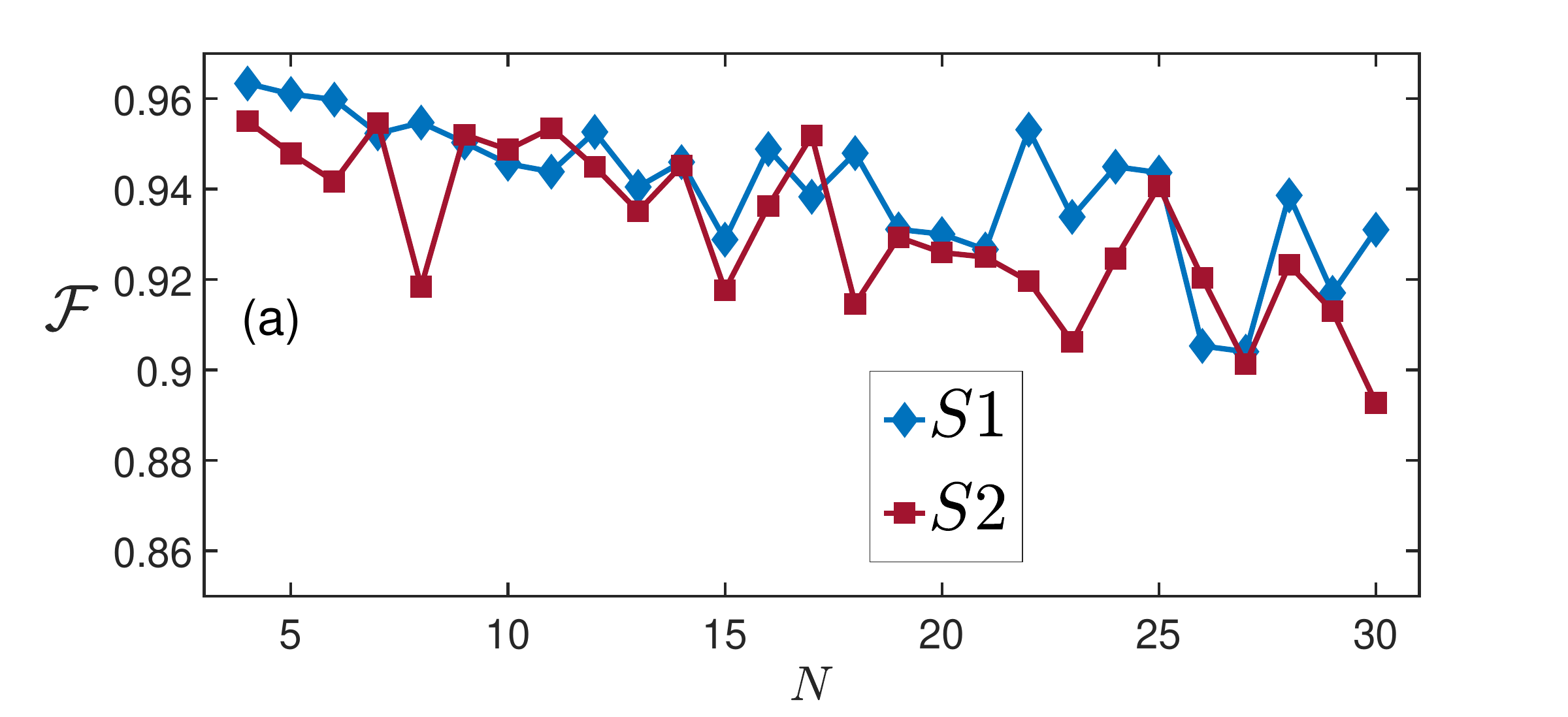}
  \quad
  \includegraphics[width=8cm,height=3.9cm]{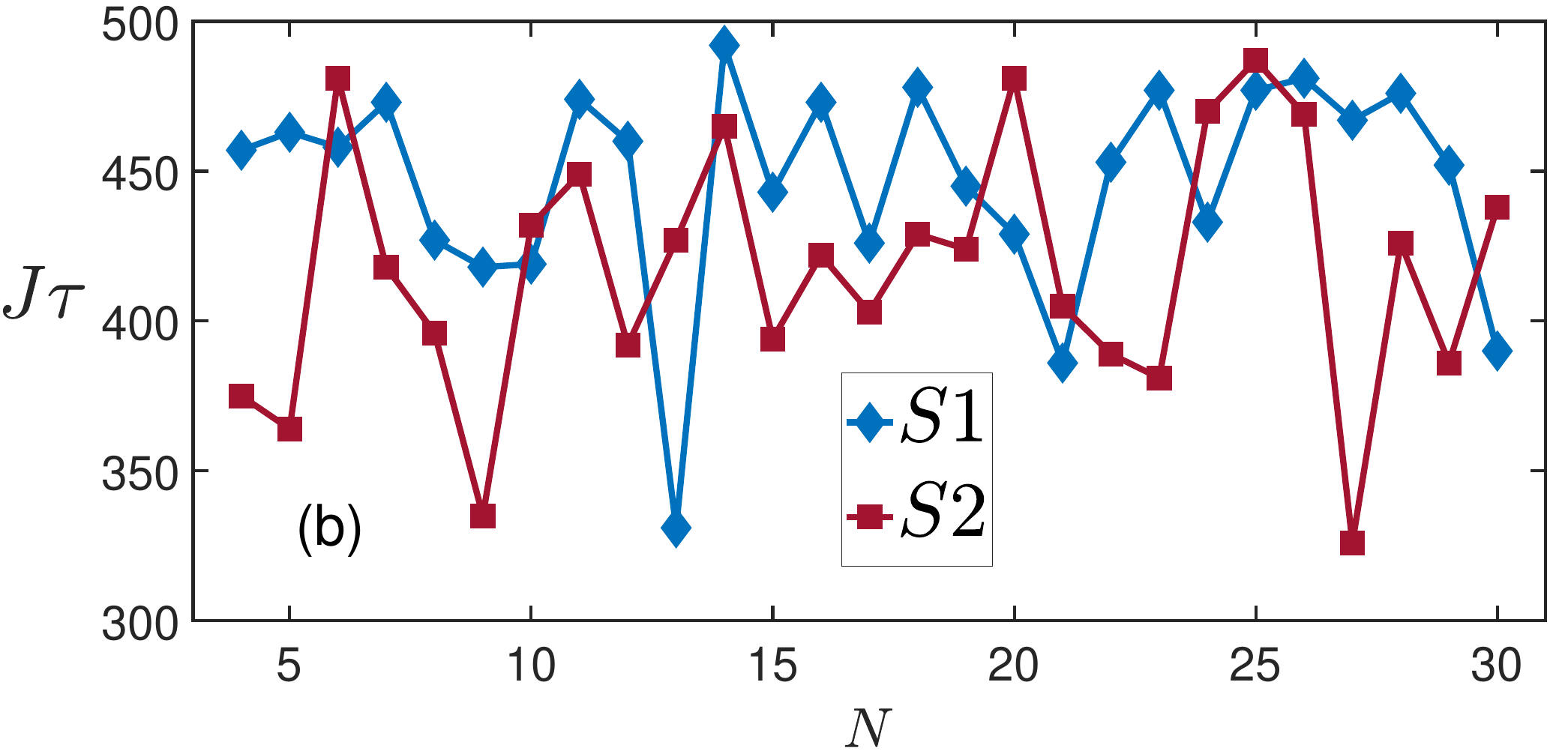}
\end{minipage}
\begin{minipage}{2\columnwidth}
\includegraphics[width=5.4cm]{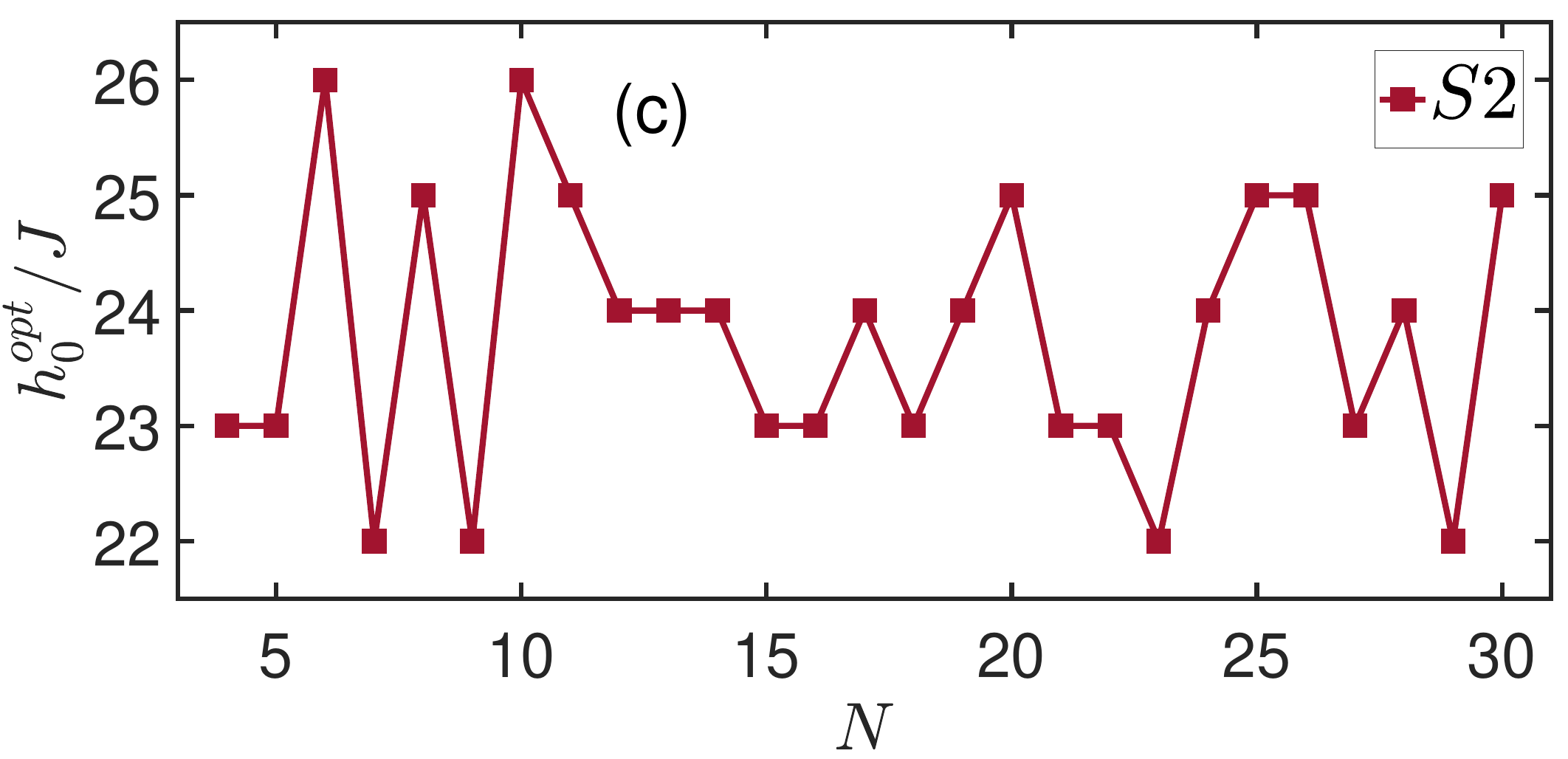}
\includegraphics[width=5.4cm]{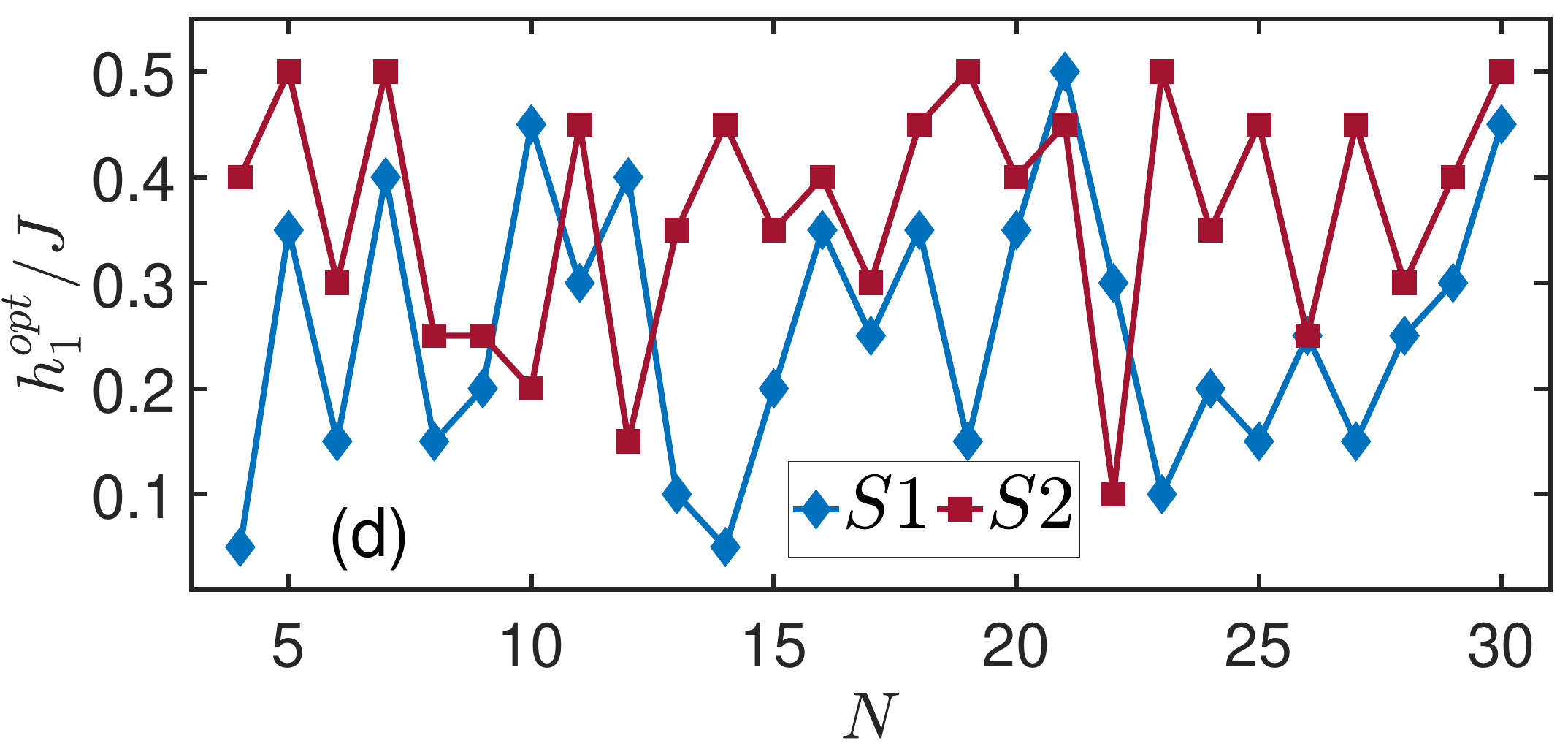}
\includegraphics[width=5.4cm]{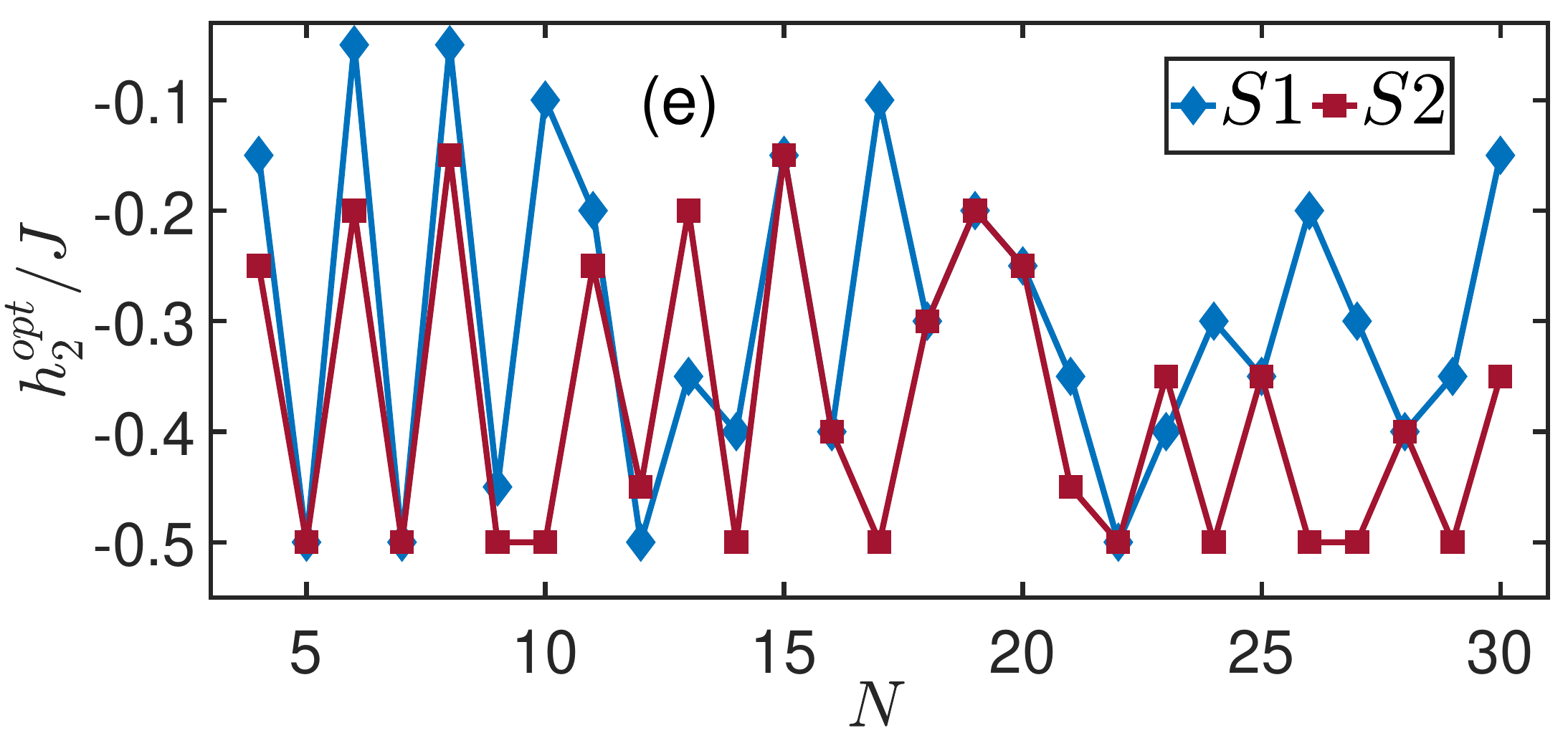}
\end{minipage}
\caption{$M=2$: (a) The scaling of optimal average gate fidelity $\mathcal{F}^{max}{=}(\overline{F}_{1}^{max}(\tau){+}\overline{F}_{2}^{max}(\tau))/2$ with $N$ for strategies $\bm{S1}$ and $\bm{S2}$. (b) The desired gate duration $\tau{\in}[1,500]/J$, (c) the optimal local magnetic field $h_0^{opt}/J{\in}[1,40]$ for establishing effective end-to-end interaction  in $\bm{S2}$, (d) and (e) the optimal magnetic fields on registers' qubits $h_{\nu}^{opt}{/}J{\in}(-1)^{\nu+1}[0,0.5]$ ($\nu=1,2$) for making the pair qubits off-resonant. }
\label{fig:FvsN}
\end{figure*}
In this section we present the numerical results for the case of two parallel gate operations.
For strategy $\bm{S1}$, the average gate fidelity $\mathcal{F}{=}(\overline{F}_{1}+\overline{F}_{2})/2$ as a function of time in chain of length $N{=}20$ is plotted in Fig.~\ref{fig:AGF vs time}(a).
Here, the Hamiltonian parameters are optimized within the chosen time interval, namely $t\in[1,500]/J$. 
The coupling $J_0$ is tuned to the optimized value $J_0^{opt}/J{=}0.04$, which results in an effective end-to-end interaction. Furthermore, we apply the optimized local magnetic fields $h_1^{opt}/J{=}0.35$ and $h_2^{opt}/J{=}{-}0.25$ on pairs $\{A_1,B_1\}$ and $\{A_2,B_2\}$, respectively, to make them energetically off-resonant and, hence, block the flow of information between them.   
For our second strategy $\bm{S2}$ the time evolution of $\mathcal{F}{=}(\overline{F}_{1}+\overline{F}_{2})/2$ is plotted in Fig.~\ref{fig:AGF vs time}(b) for a chain of length $N{=}20$ and optimized Hamiltonian parameters as   $h_0^{opt}/J{=}25$ for magnetic field applied to the end sites of the chain and $h_1^{opt}/J{=}0.4$ and $h_2^{opt}/J{=}{-}0.25$ for the magnetic fields applied on the pairs $\{A_1,B_1\}$ and $\{A_2,B_2\}$, respectively.  
As the figures show, the average gate fidelities for both strategies evolve in time and at a spatial time $t{=}\tau$ peak to their highest values which is more than $0.94$. 
In other words, by letting the system to evolve for $t=\tau$ one can perform two parallel entangling gate between the pairs $\{A_1,B_1\}$ and $\{A_2,B_2\}$, simultaneously.
 
We plot the scaling of $\mathcal{F}^{max}{=}(\overline{F}_{1}^{max}(\tau){+} \overline{F}_{2}^{max}(\tau)){/}2$ with $N$ in Fig.~\ref{fig:FvsN}(a) for our both strategies.
As the length increases the gate fidelity decreases slowly. Nonetheless, even for a pretty long chain of size $N{=}30$ the gate fidelity $\mathcal{F}^{max}$ still exceeds $0.92$. This shows the high-quality performance of parallel gate operation between two pairs of users.  
As the results illustrate, in large chains the first strategy offers better performance over the second one in terms of the gate fidelity.
The desired gate operation time $\tau$ which is optimized over the interval $t{\in}[1,500]/J$ for different $N$'s is plotted in Fig.~\ref{fig:FvsN}(b) for both $\bm{S1}$ and $\bm{S2}$. The irregular fluctuations in times is due to the fact that all the parameters are optimized for each chain and thus $\tau$ does not behave monotonically with the chain length. In particular, the most responsible parameter for the irregularities in time is $J_0$  ($h_{0}$) in strategy $\bm{S1}$ ($\bm{S2}$) which determines the effective Hamiltonian between the registers $A$ and $B$. Alternatively, one can fix theses parameters and only optimize the local fields on the registers, i.e., $h_{\nu}$, which then results more regularity in time scales~\cite{bayat2015measurement,Lorenzo2017}, though the obtained fidelities will slightly go down. Here, we give priority to fidelity instead of regularity in time. 
In the case of $\bm{S1}$, the optimal exchange coupling $J_{0}^{opt}/J{\in}[0.01,1]$ behaves  independent from the chain length and is obtained around $0.04$ for all $N$'s, consisting with the results of~\citep{yousefjani2019simultaneous,bayat2015measurement}.  
In our second strategy, the optimal magnetic fields  $h_0^{opt}$ on the end sites of the spin chain are obtained by optimizing $h_0/J{\in}[1,40]$, with only considering integer values for simplifying the optimization as the results are robust against small variation of $h_0$. The optimal results are reported in Fig.~\ref{fig:FvsN}(c) which show that for the considered chains, applying $21{<}h_{0}^{opt}/J{<}27$ is adequate for
establishing effective end-to-end interaction between registers.
The reminding Hamiltonian parameters, i.e., $h_{\nu}^{opt}{/}J{\in}(-1)^{\nu+1}[0,0.5]$ ($\nu=1,2$), are plotted in Figs.~\ref{fig:FvsN}(d) and (e) as functions of $N$. Note that the optimal magnetic fields $h_1^{opt}$ and $h_2^{opt}$ are optimized over intervals with opposite sign to increase their energy detuning.
We have used brute-force search for maximizing the fidelity when all the parameters, namely evolution time and Hamiltonian parameters, are varied over relevant intervals.
Note that the selected intervals for time evolution of the system and Hamiltonian parameters are not fundamental issues and based on the practical constraints on physical systems can be chosen differently.
To show that the intervals are not fundamental issues, later in the paper, we significantly shorten the time evolution interval which results in a much faster dynamics and still high quality gate operations. 

\subsection{Parallel gate operation for $M{>}2$}
\begin{table*}[t!]
{
\renewcommand{\arraystretch}{1.35}
\setlength{\tabcolsep}{8pt}
\begin{tabular}{| m{2mm}|  m{2cm} |  m{2cm}  m{2cm}  m{2cm}   m{1.9cm} |} 
\hline 
\multicolumn{2}{|c|}{N}& \multicolumn{1}{c}{5} & \multicolumn{1}{c}{10} & \multicolumn{1}{c}{15} & \multicolumn{1}{c|}{20} 
\\  \hline \hline
\multicolumn{1}{|c|}{\multirow{6}{*}{\rotatebox[origin=c]{90}{{$\bm{S1}$ }}}} & \multicolumn{1}{c|}{$\mathcal{F}^{max}$} & \multicolumn{1}{c}{$ 0.978$} & \multicolumn{1}{c}{$0.968$} & \multicolumn{1}{c}{$0.952$} & \multicolumn{1}{c|}{$0.947$}
\\  \cline{2-6}  
\multicolumn{1}{|c|}{}& \multicolumn{1}{c|}{$J \tau$ } & \multicolumn{1}{c}{$446$} & \multicolumn{1}{c}{$438$} & \multicolumn{1}{c}{$476$} & \multicolumn{1}{c|}{$435$}
\\  \cline{2-6}
\multicolumn{1}{|c|}{\multirow{4}{*}{}} & \multicolumn{1}{c|}{$J_{0}^{opt}/J$}  & \multicolumn{1}{c}{$0.04$} & \multicolumn{1}{c}{$0.04$} & \multicolumn{1}{c}{$0.04$} & \multicolumn{1}{c|}{$0.04$}
\\  \cline{2-6}
\multicolumn{1}{|c|}{\multirow{3}{*}{}} & \multicolumn{1}{c|}{$h_{1}^{opt}/J$}  & \multicolumn{1}{c}{$0.4$} & \multicolumn{1}{c}{$0.5$} & \multicolumn{1}{c}{$0.2$} & \multicolumn{1}{c|}{$0.35$}
\\  \cline{2-6} 
\multicolumn{1}{|c|}{}& \multicolumn{1}{c|}{$h_{2}^{opt}/J$} & \multicolumn{1}{c}{$-0.3$} & \multicolumn{1}{c}{$-0.1$} & \multicolumn{1}{c}{$-1.2$} & \multicolumn{1}{c|}{$-0.25$}
\\  \cline{2-6} 
\multicolumn{1}{|c|}{}& \multicolumn{1}{c|}{$h_{3}^{opt}/J$ } & \multicolumn{1}{c}{$0.35$} & \multicolumn{1}{c}{$0.4$} & \multicolumn{1}{c}{$0.6$} & \multicolumn{1}{c|}{$0.05$}                       
\\  \hline 
\end{tabular}
\qquad
\begin{tabular}{| m{2mm}|  m{2cm} |  m{2cm}  m{2cm}  m{2cm}   m{1.9cm} |} 
\hline 
\multicolumn{2}{|c|}{N}& \multicolumn{1}{c}{5} & \multicolumn{1}{c}{10} & \multicolumn{1}{c}{15} & \multicolumn{1}{c|}{20} 
\\  \hline \hline
\multicolumn{1}{|c|}{\multirow{6}{*}{\rotatebox[origin=c]{90}{{$\bm{S2}$}}}} & \multicolumn{1}{c|}{$\mathcal{F}^{max}$} & \multicolumn{1}{c}{$0.977$} & \multicolumn{1}{c}{$0.963$} & \multicolumn{1}{c}{$0.947$} & \multicolumn{1}{c|}{$0.919$}
\\  \cline{2-6}  
\multicolumn{1}{|c|}{}& \multicolumn{1}{c|}{$J \tau$ } & \multicolumn{1}{c}{$459$} & \multicolumn{1}{c}{$472$} & \multicolumn{1}{c}{$482$} & \multicolumn{1}{c|}{$500$}
\\  \cline{2-6}
\multicolumn{1}{|c|}{\multirow{4}{*}{}} & \multicolumn{1}{c|}{$h_{0}^{opt}/J$}  & \multicolumn{1}{c}{$26$} & \multicolumn{1}{c}{$25$} & \multicolumn{1}{c}{$26$} & \multicolumn{1}{c|}{$28$}
\\  \cline{2-6}
\multicolumn{1}{|c|}{\multirow{3}{*}{}} & \multicolumn{1}{c|}{$h_{1}^{opt}/J$}  & \multicolumn{1}{c}{$0.5$} & \multicolumn{1}{c}{$0$} & \multicolumn{1}{c}{$0.2$} & \multicolumn{1}{c|}{$1$}
\\  \cline{2-6} 
\multicolumn{1}{|c|}{}& \multicolumn{1}{c|}{$h_{2}^{opt}/J$} & \multicolumn{1}{c}{$-1.1$} & \multicolumn{1}{c}{$-0.7$} & \multicolumn{1}{c}{$-0.7$} & \multicolumn{1}{c|}{$-0.6$}
\\  \cline{2-6} 
\multicolumn{1}{|c|}{}& \multicolumn{1}{c|}{$h_{3}^{opt}/J$ } & \multicolumn{1}{c}{$1.1$} & \multicolumn{1}{c}{$1.2$} & \multicolumn{1}{c}{$1$} & \multicolumn{1}{c|}{$1.2$}                       
\\  \hline 
\end{tabular}
\caption{$M{=3}$: The maximum of $\mathcal{F}$ for gate duration $\tau\in[1,500]/J$ by adopting strategies $1$ and $2$ in different chains. Here, the optimal exchange coupling $J_0^{opt}/J$ for strategy $\bm{S1}$ has been optimized over the interval $J_0^{opt}/J{\in}[0.01,1]$ and the optimal local magnetic field on the ends of the chain $h_0^{opt}/J$ for $\bm{S2}$ has been optimized over $h_0^{opt}/J{\in}[1,40]$. In both strategies, the optimal values for the local fields on qubits of the registers, i.e., $h_1^{opt}/J$, $h_2^{opt}/J$ and $h_3^{opt}/J$, have been optimized over the interval $(-1)^{\nu+1}[0,1.5]$.}\label{table1}
}
\end{table*}
\begin{table*}[t!]
{
\renewcommand{\arraystretch}{1.25}
\setlength{\tabcolsep}{8pt}
\begin{tabular}{| m{2mm}|  m{2cm} |  m{2cm}  m{2cm}  m{2cm}   m{1.9cm} |} 
\hline 
\multicolumn{2}{|c|}{M}& \multicolumn{1}{c}{1} & \multicolumn{1}{c}{2} & \multicolumn{1}{c}{3} & \multicolumn{1}{c|}{4} 
\\  \hline \hline
\multicolumn{1}{|c|}{\multirow{7}{*}{\rotatebox[origin=c]{90}{{$\bm{S1}$ }}}} & \multicolumn{1}{c|}{$\mathcal{F}^{max}$} & \multicolumn{1}{c}{$ 0.987$} & \multicolumn{1}{c}{$0.968$} & \multicolumn{1}{c}{$0.966$} & \multicolumn{1}{c|}{$0.953$}
\\  \cline{2-6}  
\multicolumn{1}{|c|}{}& \multicolumn{1}{c|}{$J \tau$ } & \multicolumn{1}{c}{$53$} & \multicolumn{1}{c}{$45$} & \multicolumn{1}{c}{$97$} & \multicolumn{1}{c|}{$81$}                     
\\  \cline{2-6}  
\multicolumn{1}{|c|}{}& \multicolumn{1}{c|}{$J_{0}^{opt}/J$ } & \multicolumn{1}{c}{$0.06$} & \multicolumn{1}{c}{$0.13$} & \multicolumn{1}{c}{$0.09$} & \multicolumn{1}{c|}{$0.11$}                     
\\  \cline{2-6}  
\multicolumn{1}{|c|}{}& \multicolumn{1}{c|}{$h_{1}^{opt}/J$ } & \multicolumn{1}{c}{$0.68$} & \multicolumn{1}{c}{$0.24$} & \multicolumn{1}{c}{$1.02$} & \multicolumn{1}{c|}{$1.4$}
\\  \cline{2-6}  
\multicolumn{1}{|c|}{}& \multicolumn{1}{c|}{$h_{2}^{opt}/J$ } & \multicolumn{1}{c}{$-$} & \multicolumn{1}{c}{$-1$} & \multicolumn{1}{c}{$-1.4$} & \multicolumn{1}{c|}{$-1.1$}                                          
\\  \cline{2-6}  
\multicolumn{1}{|c|}{}& \multicolumn{1}{c|}{$h_{3}^{opt}/J$ } & \multicolumn{1}{c}{$-$} & \multicolumn{1}{c}{$-$} & \multicolumn{1}{c}{$0.14$} & \multicolumn{1}{c|}{$0.05$}                     
\\  \cline{2-6}  
\multicolumn{1}{|c|}{}& \multicolumn{1}{c|}{$h_{4}^{opt}/J$ } & \multicolumn{1}{c}{$-$} & \multicolumn{1}{c}{$-$} & \multicolumn{1}{c}{$-$} & \multicolumn{1}{c|}{$-1.3$}                     
\\  \hline 
\end{tabular}
\qquad
\begin{tabular}{| m{2mm}|  m{2cm} |  m{2cm}  m{2cm}  m{2cm}   m{1.9cm} |} 
\hline 
\multicolumn{2}{|c|}{M}& \multicolumn{1}{c}{1} & \multicolumn{1}{c}{2} & \multicolumn{1}{c}{3} & \multicolumn{1}{c|}{4} 
\\  \hline \hline
\multicolumn{1}{|c|}{\multirow{7}{*}{\rotatebox[origin=c]{90}{{$\bm{S2}$}}}} & \multicolumn{1}{c|}{$\mathcal{F}^{max}$} & \multicolumn{1}{c}{$0.981$} & \multicolumn{1}{c}{$0.952$} & \multicolumn{1}{c}{$0.928$} & \multicolumn{1}{c|}{$0.924$}
\\  \cline{2-6}  
\multicolumn{1}{|c|}{}& \multicolumn{1}{c|}{$J \tau$ } & \multicolumn{1}{c}{$80$} & \multicolumn{1}{c}{$90$} & \multicolumn{1}{c}{$98$} & \multicolumn{1}{c|}{$88$}                     
\\  \cline{2-6}  
\multicolumn{1}{|c|}{}& \multicolumn{1}{c|}{$h_{0}^{opt}/J$ } & \multicolumn{1}{c}{$10$} & \multicolumn{1}{c}{$14$} & \multicolumn{1}{c}{$12$} & \multicolumn{1}{c|}{$12$}                     
\\  \cline{2-6}  
\multicolumn{1}{|c|}{}& \multicolumn{1}{c|}{$h_{1}^{opt}/J$ } & \multicolumn{1}{c}{$0.15$} & \multicolumn{1}{c}{$0.59$} & \multicolumn{1}{c}{$1.46$} & \multicolumn{1}{c|}{$0.45$}
\\  \cline{2-6}  
\multicolumn{1}{|c|}{}& \multicolumn{1}{c|}{$h_{2}^{opt}/J$ } & \multicolumn{1}{c}{$-$} & \multicolumn{1}{c}{$-0.71$} & \multicolumn{1}{c}{$-1.28$} & \multicolumn{1}{c|}{$-0.3$}                                          
\\  \cline{2-6}  
\multicolumn{1}{|c|}{}& \multicolumn{1}{c|}{$h_{3}^{opt}/J$ } & \multicolumn{1}{c}{$-$} & \multicolumn{1}{c}{$-$} & \multicolumn{1}{c}{$0.16$} & \multicolumn{1}{c|}{$1.3$}                     
\\  \cline{2-6}  
\multicolumn{1}{|c|}{}& \multicolumn{1}{c|}{$h_{4}^{opt}/J$ } & \multicolumn{1}{c}{$-$} & \multicolumn{1}{c}{$-$} & \multicolumn{1}{c}{$-$} & \multicolumn{1}{c|}{$-1.4$}                     
\\  \hline 
\end{tabular}
\caption{Comparison: The maximum of $\mathcal{F}$ and desired gate duration $\tau\in[1,100]/J$ for different number of pairs $M=1,\ldots,4$ in a chain with $N=4$ by adopting outlined strategies. In the case of $\bm{S1}$, the optimal exchange coupling, $J_{0}^{opt}/J$, is obtained by surfing on the interval $[0.01,1]$. The optimal magnetic field, $h_{0}^{opt}/J$, is optimized over $[1,40]$ for the case of $\bm{S2}$. In both strategies and for all $M$'s the optimal local magnetic fields on the registers' qubits, $h_{\nu}^{opt}/J$ ($\nu=1,\ldots,4$), are optimized over $(-1)^{\nu+1}[0,1.5]$. }\label{table2}
}
\end{table*}
In this section we show that the parallel gate operation can be extended beyond $M{=}2$. In fact, arbitrary number of parallel gates can be performed using our outlined strategies. In TABLE ~\ref{table1}, we present the performance of our protocol for the case of $M{=}3$ by adopting two strategies $\bm{S1}$ and $\bm{S2}$ for different chains. Here, $\mathcal{F}^{max}=(\overline{F}_{1}^{max}(\tau)+\overline{F}_{2}^{max}(\tau)+\overline{F}_{3}^{max}(\tau))/3$ is obtained after embedding the optimal values of the Hamiltonian parameters, i.e., $J_{0}^{opt}{/}J{\in}[0.01,1]$, $h_{0}^{opt}{/}J{\in}[20,40]$,
and $h_{\nu}^{opt}{/}J{\in}(-1)^{\nu+1}[0,1.5]$ ($\nu=1,2,3$), presented in TABLE~\ref{table1}, which are obtained within the time window $t{\in}[1,500]/J$. 
As results show, regardless of the adopted strategy, the gate fidelity achieves very high values such that $\mathcal{F}^{max}$ remains larger than $0.91$ even for chains up to $N{=}20$. Similar to the case of $M{=}2$, the first strategy presents better performance than the other one for long chains.

\section{Advantages of the parallelism}
To highlight the advantages offered by our parallel gate operation protocol, in TABLE~\ref{table2} we report $\mathcal{F}^{max}$ and the optimal time $\tau$ for simultaneously implanting $M{=}1,{\ldots},4$ entangling gates across a chain of length $N{=}4$.
On the contrary to the previous sections, here, we restrict the time interval to $t{\in}[1,100]/J$ and find the optimal time $\tau$ over this time period.
Interestingly, for both strategies, the average gate fidelities are steadily high and a comparison between the desired gate durations, $\tau$, for achieving this fidelities shows that the operation rate of our protocol for $M{>}1$ is always faster than the $M$ sequential implementations of the gate.
To have a quantitative analysis, we focus on the case of $M{=}4$.  In the sequential strategy, one has to 
operate the gate $M$ times while in the parallel protocol all the $M$ gates are performed simultaneously. Using the data shown in TABLE~\ref{table2}, we can see that in the sequential strategy one has to apply $G_1$ for $M{=}4$ subsequent times which results in the total time of $\tau_{\text{seq}}{=}4 {\times} 53 {=} 212$ for $\bm{S1}$ and $\tau_{\text{seq}}{=}4 {\times}80{=}320$ for $\bm{S2}$. While the parallel protocol only demands $\tau_{\text{par}}{=}81$ for $\bm{S1}$ and $\tau_{\text{par}}{=}88$ for $\bm{S2}$ through a single operation.  
This shows that, in the strategy $\bm{S2}$ ($\bm{S1}$), the parallel implementation of the entangling gates is $3.6$ ($2.6$) times faster than the sequential approach.
These results indicate that, irrespective of the adapted strategy, our parallel protocol remarkably accelerates the rate of implementing two-qubit gates with no extra cost.

It is worth emphasizing that, in general, entangling gate operations across spin chains can be classified into two major groups. In the first group, relying on linear dispersion relation~\cite{christandl2004perfect,christandl2005perfect,di2008perfect,kay2006perfect}, one may use the coherent wave packet propagation mechanism to establish entangling gate operations. 
While resorting to this mechanism provides fast and reusable entangling gates for $M{=}1$ pair of qubits~\cite{banchi2011nonperturbative}, parallelism is hard to achieve   
as many eigenstates involve in the evolution.
In the second group, the data-bus is used as an interaction mediator that only virtually populated~\cite{bayat2015measurement} and, hence, the
dynamics is governed by an effective Hamiltonian between distant qubits which are off-resonant from the channel by either using weak couplings or strong local magnetic fields.
The reduction to an effective few-qubit system usually comes at the price of increasing the time operations~\cite{yao2011robust}. However, our protocol shows that by adopting parallel gate operation one can significantly reduce the time in the latter group allowing for rapid parallel gate operations.  

\section{PERFORMANCE UNDER REALISTIC CONDITIONS}
Any physical realization of entangling gates will inevitably deviate from the ideal scenario due to the disorders.  
The source of these disorders can be imperfections in the fabrication or decoherence due to interaction with an environment.
This section is dedicated to analyze the effect of these two types of disorders on the performance of our protocol.
Here, without loss of generality, we focus on the case $M{=}2$.
\subsection{Fabrication Imperfections}
\begin{figure}[t!]
	\centering\offinterlineskip
	\includegraphics[width=\linewidth]{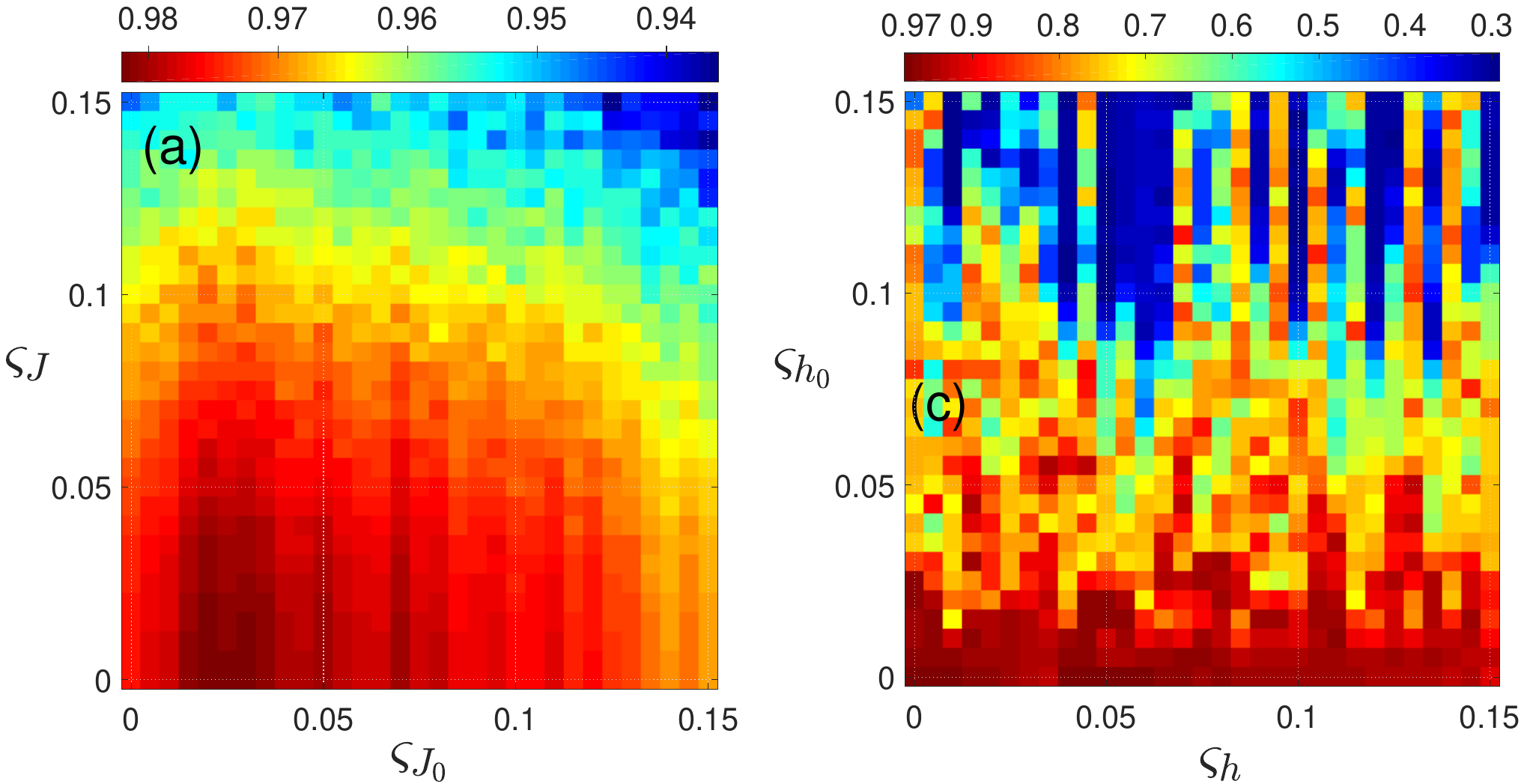}
	\includegraphics[width=\linewidth]{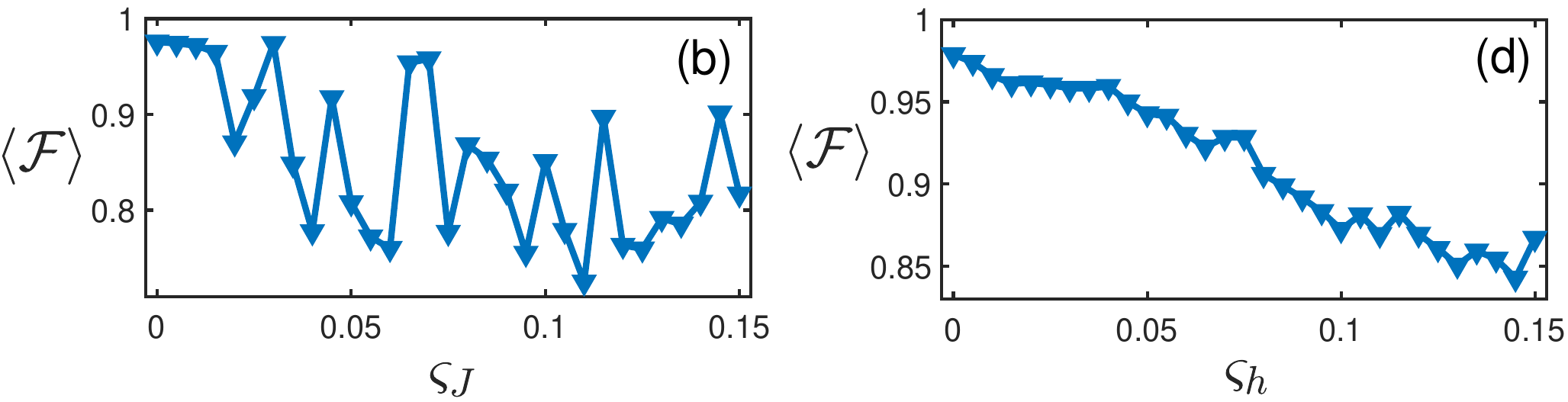}
	\caption{Fabrication imperfections: (a) ((c)) Average gate fidelity $\langle \mathcal{F} \rangle$  in term of $\varsigma_{J} , \varsigma_{J_{0}}$ ($\varsigma_{h_{0}} , \varsigma_{h}$) in spin-chain channel with length $N{=}10$ for strategy $\bm{S1}$ ($\bm{S2}$). In (b) and (c), $\langle \mathcal{F} \rangle$ as functions of $\varsigma_{J}$ and $\varsigma_{h}$ is depicted in the same channel for the second and first strategies, respectively. The Hamiltonian parameters for obtaining these plots are considered as $\{J_0/J{=}0.04, h_1/J{=}0.45, h_2/J{=}{-}0.1\}$ and $\{h_0/J{=}26, h_1/J{=}0.2, h_2/J{=}{-}0.5\}$, respectively, for $\bm{S1}$ and $\bm{S2}$. In all the plots, $\langle \mathcal{F} \rangle$ is calculated as the ensemble average of $100$ different realizations of the gate implementation for certain $\varsigma \in \{\varsigma_{J},\varsigma_{J_{0}},\varsigma_{h_{0}},\varsigma_{h}\}$.    }\label{fig:Imperfection}
\end{figure}

In the previous sections, we established parallel entangling gate operations between distant pairs of qubits  just by tuning Hamiltonian parameters appropriately.  
In practice due to the imperfections, there is no guarantee that the Hamiltonian parameters can be tuned perfectly. 
In particular, the exchange couplings and the local magnetic fields are likely to be
subject to random variations and the actual values of them may not be precisely known, leading to an uncertainly in the system Hamiltonian.
For the sake of clarity, here, the impacts of randomization in the exchange couplings and the local magnetic fields would be analyzed  individually.  
To figure out how small random variations in the exchange couplings are likely to affect the effectiveness of our protocol, we consider the coupling between neighboring sites $(i,i+1)$ inside the channel as $J(1+\chi^{J}_{i})$ and the coupling between the registers and the channel as $J_{0}(1+\chi^{J_{0}}_{\nu})$. Here, $\chi^{J}_{i}$ and $\chi^{J_{0}}_{\nu}$ are randomly sampled from a normal probability distribution with zero mean and variance $\varsigma_{J}$ and $\varsigma_{J_{0}}$, respectively.    
To assess the effectiveness of randomization in the local magnetic fields, we assume that the magnitude of applied fields on the end sites of the channel and registers' qubits vary as $h_{0}(1+\chi^{h_{0}})$ and $h_{\nu}(1+\chi^{h_{\nu}}_{\nu})$, where $\chi^{h_{0}}$ and $\chi^{h_{\nu}}_{\nu}$ are again uniformly distributed random variable with mean $0$ and variance $\varsigma_{h_{0}}$ and $\varsigma_{h}$, respectively.  
In order to see the effect of disorder on the quality of gate operation for each certain magnitude of variance $\varsigma \in \{\varsigma_{J},\varsigma_{J_{0}},\varsigma_{h_{0}},\varsigma_{h}\}$, we create $100$ different random instances and then compute the ensemble average of attainable gate fidelity $\langle \mathcal{F} \rangle$.
In Fig.~\ref{fig:Imperfection}(a) the random averaged gate fidelity $\langle \mathcal{F} \rangle$  is depicted as a function of $\varsigma_{J}$ and $\varsigma_{J_{0}}$ in a spin-chain of $N{=}10$ when the first strategy $\bm{S1}$ is adopted. 
The figure shows that although the susceptibility of our first strategy to the uncertainty and inhomogeneity in $J$, is more than that in $J_{0}$, the protocol efficiency remains stable for variations up to $10 \%$ of the exchange couplings.
This stability is a direct consequence of preserving the condition $J_{0}{\ll}J$ which guarantees that the channel mediates an effective interaction between the registers. 
Fig.~\ref{fig:Imperfection}(b) illustrates $\langle \mathcal{F} \rangle$ as a function of $\varsigma_{J}$ in the same spin-chain channel by adopting strategy $\bm{S2}$.    
A comparison between the efficiency of our two strategies exhibits that $\bm{S1}$ is more robust than $\bm{S2}$ to randomness in exchange couplings. 
In Fig.~\ref{fig:Imperfection}(c) the random averaged fidelity $\langle \mathcal{F} \rangle$ of our second strategy is plotted as a function of $\varsigma_{h_{0}}$ and $\varsigma_{h}$ in a chain of length $N{=}10$. While the protocol rapidly loses its efficiency by increasing the uncertainty in the applied magnetic field on the end sits of the channel, it shows more robust behavior against the randomness in $h_{\nu}$s. The results show that disorder in $h_{0}$ ($h_{\nu}$) up to $2.5 \%$ ($15 \%$) of the main value of the parameter is tolerable.   
Finally, $\langle \mathcal{F} \rangle$ as a function of $\varsigma_{h}$ for strategy $\bm{S1}$ in the same spin-chain channel is plotted in Fig.~\ref{fig:Imperfection}(d). Obviously, the efficiency of the protocol decreases gradually while the uncertainty in applied magnetic fields on registers' qubits is enhanced. Nonetheless, for disorder up to $7.5 \%$ of $h_{\nu}$'s, one reaches $\langle \mathcal{F} \rangle{>}0.9$.
Note that, in preparing Fig.~\ref{fig:Imperfection}, the Hamiltonian parameters for each strategy are set in a way that $\langle\mathcal{F}\rangle{=}\mathcal{F}^{max}$ for vanishing disorders.
A total comparison of the results show that, in general, our first strategy is more robust against fabrication imperfections than the second one. 

\subsection{Decoherence}
\begin{figure}[t!]
	\centering\offinterlineskip
	\includegraphics[width=\linewidth]{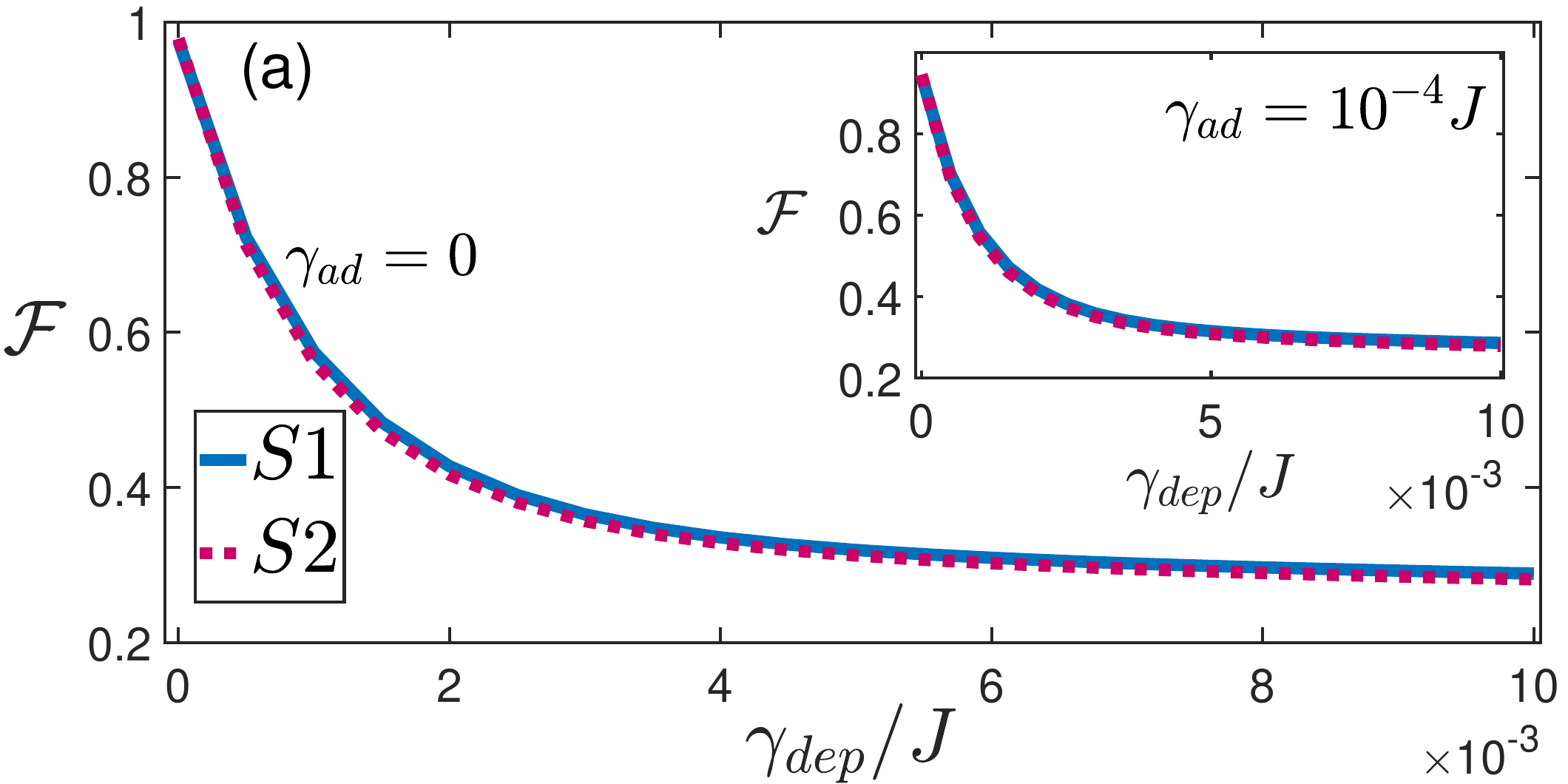}
		\includegraphics[width=\linewidth]{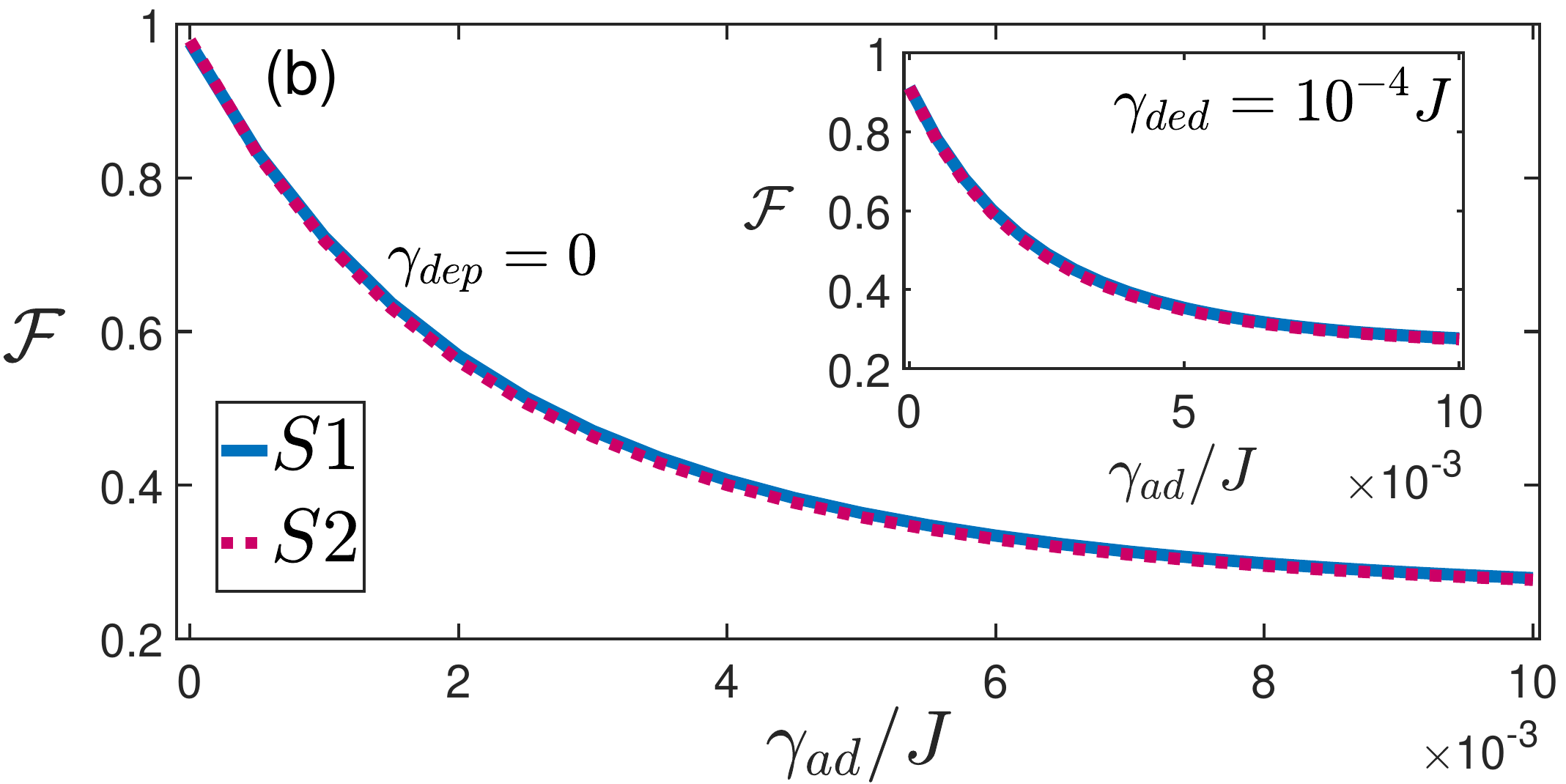}
	\caption{Decoherence: The average $\mathcal{F}{=}(\overline{F}_{1}+\overline{F}_{2})/2$ for our two strategies $\bm{S1}$ and $\bm{S2}$ for pure dephasing noise as function of $\gamma_{dep}/J$ (a) and pure amplitude damping noise as function of $\gamma_{ad}/J$ (b) in a chain of $N{=}10$. Insets of (a) and (b) show the average gate fidelities for impure phase and amplitude damping noises as functions of one noise strength while the other one is fixed. 
The Hamiltonian parameters  are taken as $\{J_0/J{=}0.04, h_1/J{=}0.45, h_2/J{=}{-}0.1\}$ and $\{h_0/J{=}26, h_1/J{=}0.2, h_2/J{=}{-}0.5\}$, respectively, for $\bm{S1}$ and $\bm{S2}$.}\label{fig:Dephasing}
\end{figure}

In practice, the deteriorative effects of quantum noises resulting from interactions with the environment are inevitable. 
In this section, we analyze the resistance of our protocol against two types of serious noises in the entangling gates, including dephasing and amplitude damping. 
Dephasing arises from fluctuating magnetic or electric fields in the environment and while it changes the quantum superposition to a classical mixtures, it conserves the number of excitations. On the other hand, the amplitude damping noise involves exchanging excitations with the environment and thus the number of excitations does not conserve. 
Here, we assume that the interaction of the system and environment is Markovian type and acts simultaneously with turning on the coupling $J_{0}$, therefore, the evolution of the system can be described by a Lindblad master equation  
\begin{eqnarray}\label{master}
\dfrac{d\rho(t)}{dt}&=&-i[H,\rho(t)]+ \gamma_{dep}\sum_{i} \left(\sigma_i^z \rho(t)\sigma_{i}^z{-}\rho(t)\right) \nonumber \\ 
&+& \gamma_{ad}\sum_{i} \left(\sigma_i^- \rho(t) \sigma_i^+ -\dfrac{1}{2} \{\sigma_i^+\sigma_i^- ,\rho(t)\} \right)\nonumber \\
\end{eqnarray}
where $\gamma_{dep}$ ($\gamma_{ad}$) represents the strength of the dephasing (amplitude damping) noise, $\sigma_i^{\pm}{=}(\sigma_i^{x}\pm i\sigma_i^{y})/2$ and $\{,\}$ denotes anticommutator.  
Here, we consider the performance of our two strategies $\bm{S1}$ and $\bm{S2}$ in a spin-chain of $N{=}10$ against the increase of the noises strength. 
The behaviors of the gate fidelity versus  pure dephasing ($\gamma_{ad}{=}0$) and pure amplitude damping ($\gamma_{dep}{=}0$) are plotted in Figs.~\ref{fig:Dephasing}(a) and (b), respectively.    
To see the effect of both noises together, in the insets of Figs.~\ref{fig:Dephasing}(a) and (b) we plot the fidelity  $\mathcal{F}$ as a function of one noise strength while keeping the other is fixed. 
The Hamiltonian parameters are optimized for the noise free system. As expected the performance goes down as the noise strength increases. In addition, both of the strategies operate with the same quality, regardless of the noise type.
In a practical scheme, if one aims for the fidelity to be $\mathcal{F}{>}0.9$,  one has to fabricate the device such that $\gamma_{dep}/J \sim 0.01$ and $\gamma_{ad}/J\sim 0.05 $, respectively.
 
\section{Experimental proposal}
Superconducting transmonic devices are the leading platforms for quantum simulations~\cite{Roushan2017,Yan2019,gong2019genuine,guo2020observation}.
In this section, we show that our protocol can indeed be realized in such set-ups. In a typical coupled superconducting qubits, the exchange coupling strengths between nearest  neighbors  can be tuned up to $J{\simeq} 50$~MHz and independent control on each qubit allows local energy splitting between $0{-}800$~MHz~\cite{Yan2019,Roushan2017}.
Implementing parallel two-qubit gates on $M{=}3$ distant pairs of superconducting qubits that are coupled to a data-bus integrated of $N{=}4$ transmon qubits with the identical coupling strength $J{\simeq} 50$~MHz (strategy $\bm{S2}$) results in $\mathcal{F}{>}0.92$ at $\tau{=}1.96$~$\mu$s. 
For obtaining such average gate fidelity, one needs to properly set the Hamiltonian parameters as $h_{0}{=}600$~MHz (${=}12J$), $h_{1}{=}73$~MHz (${=}1.46J$), $h_{2}{=}{-}64$~MHz (${=}{-}1.28J$), and $h_{3}{=}8$~MHz (${=}0.16J$). For these choices of parameters, the evolution time window will be $\tau {\in} [0,10]$~$\mu$s. 
The relaxation and dephasing times of such devises are $T_{1}{\simeq}50{-}85$~$\mu$s and $T_{2}{\simeq}10{-}50$~$\mu$s~\cite{Kjaergaard2020}, respectively. Thus, one can estimate the lowest values of noise strength as 
$\gamma_{dep} {\simeq} 20$~KHz (${=}4\times 10^{-4}J$) and $\gamma_{ad}{\sim} 11$~KHz (${=}2.2\times 10^{-4}J$). Surprisingly, for these values of noise strength, the attainable average gate fidelity is obtained as $\mathcal{F}{>}0.74$. By improving the coherence of the system or enhancing the couplings between the qubits the fidelity can be improved even further.
\section{Two-way quantum communication}\label{se:Two-way}
In Ref.~\cite{PhysRevA.84.022345}, the authors propose a two-way quantum
communication setup in which two users can exchange quantum states at the same type using the same spin chain data-bus, with very low fidelity. Our protocol for implementing parallel two-qubit gates  can also be used for high fidelity two-way quantum communication between the registers $A$ and $B$. Indeed, the conditions (I) and (II) are adequate to construct a high-fidelity two-way quantum communication.
For the case of $M{=2}$, consider the users $\{A_1,B_2\}$ ($\{B_1,A_2\}$) as senders (receives). 
To establish our two-way communication, we assume the total initial state Eq.~(\ref{eq:Initial state}) as $\vert \Psi_{0}\rangle=\vert \psi_{1},\psi_{2}\rangle_{A}|\bm{0}\rangle_{ch}\vert \varphi_{1},\varphi_{2}\rangle_{B}$ with   
$|\psi_{1}\rangle_{A}{=}\cos(\theta_{1}/2)|0\rangle + e^{i\phi_{1}}\sin(\theta_{1}/2)|1\rangle $, $|\psi_{2}\rangle_{A}{=}|0\rangle$, $|\varphi_{1}\rangle_{B}{=}|0\rangle$, and
$|\varphi_{2}\rangle_{B}{=}\cos(\theta_{2}/2)|0\rangle + e^{i\phi_{2}}\sin(\theta_{2}/2)|1\rangle$.
As each receiver, namely $B_{1}$ and $A_{2}$, accepts the information that arrives from both senders, to quantify the performance of our protocol we define transmission fidelity 
$F_T(t){=}\vert \langle \Phi_{T} \vert e^{-iHt} \vert \Psi_{0} \rangle \vert^{2}$, with $\vert \Phi_{T} \rangle {=}|\varphi_{1},\varphi_{2}\rangle_{A}  |\bm{0}\rangle_{ch}|\psi_{1},\psi_{2}\rangle_{B}$ and crosstalk 
$F_C(t){=}\vert \langle \Phi_{C} \vert e^{-iHt} \vert \Psi_{0} \rangle \vert^{2}$, with $\vert \Phi_{C} \rangle {=}|\psi_{2},\psi_{1}\rangle_{A}  |\bm{0}\rangle_{ch}|\varphi_{2},\varphi_{1}\rangle_{B}$, caused by the information flow between $\{A_1,B_1\}$ and $\{A_2,B_2\}$.
By taking the average of these quantities over all possible initial states on the surface of the Bloch spheres, one gets the input-independent quantities
$\bar{F}_{T,C}(t){=}\int d\Omega_{1} d\Omega_{2} F_{T,C}(t)$, where $d\Omega_{\nu}=\dfrac{1}{4\pi}\sin(\theta_{\nu})d\theta_{\nu} d\phi_{\nu}$, with $\nu{=}1,2$, is the normalized SU(2) Haar measure.
For the sake of completeness, in the following we   
first investigate the fidelity for different states, helping to know the
performance of the protocol in the worst scenario,  namely the minimal attainable fidelity. Next, we analyze the average transmission fidelity and corresponding crosstalk.

\begin{figure}[t!]
	\centering\offinterlineskip
	\includegraphics[width=\linewidth]{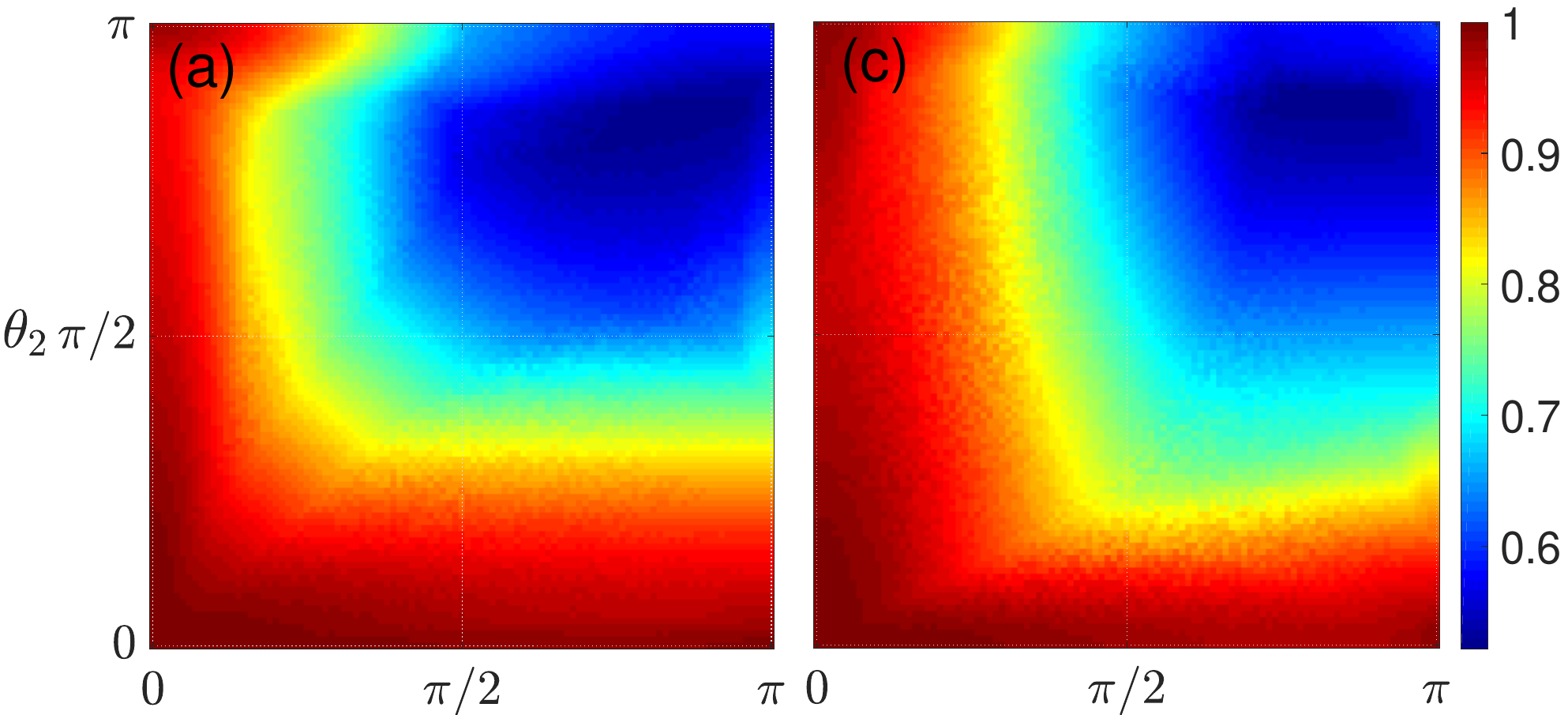}
	\includegraphics[width=\linewidth]{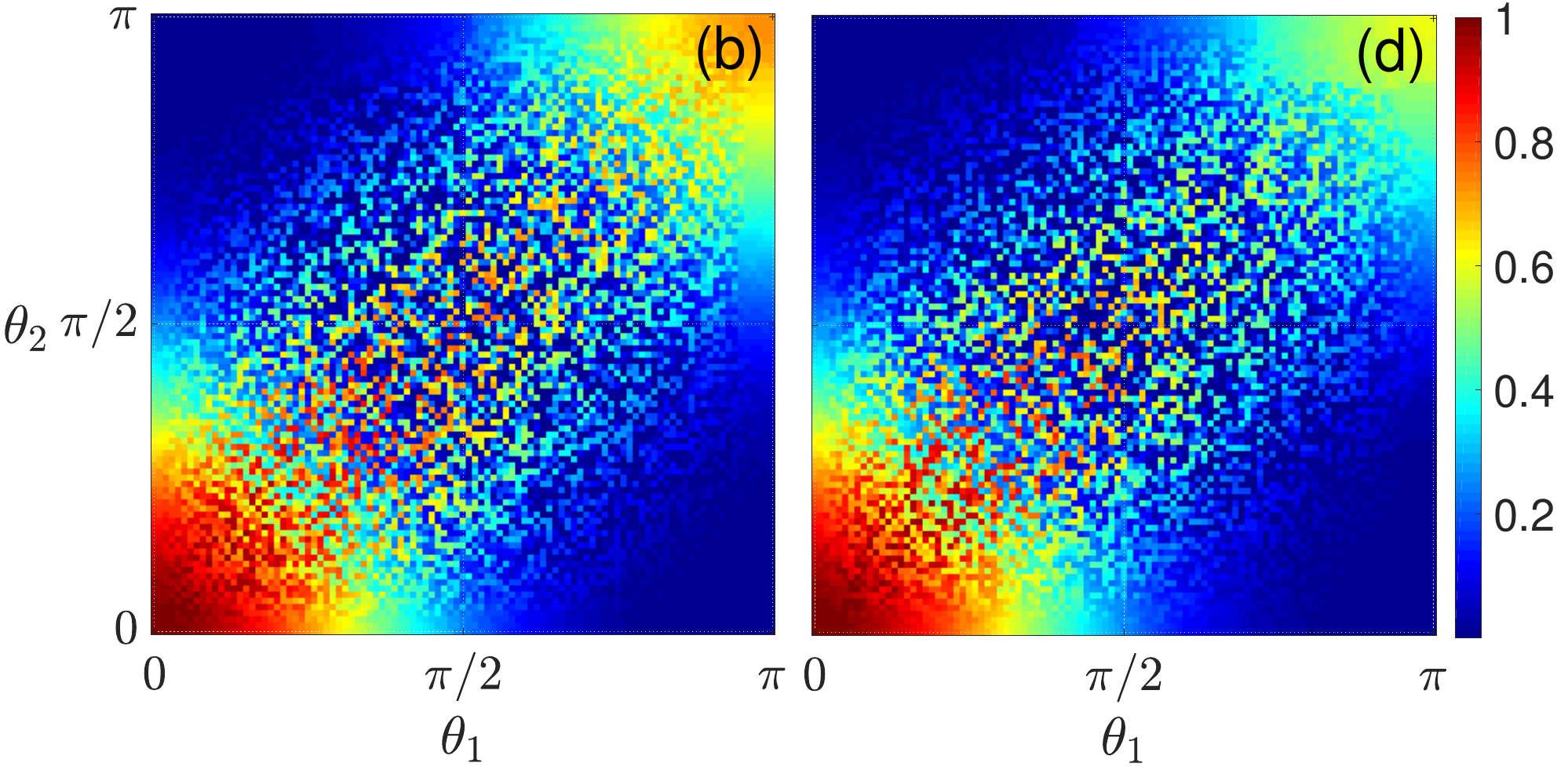}
	\caption{Two-way communication: The maximal transmission fidelity $F_{T}^{max}(\tau)$ (a) ((c)) and the corresponding crosstalk $F_{c}(\tau)$ (b) ((d))  for strategy $\bm{S1}$ ($\bm{S2}$) as a function of $\theta_{1}$ and $\theta_{2}$ in a chain of $N{=}10$ and at optimal arriving time $\tau$. The Hamiltonian parameters are taken as $\{J_{0}^{opt}/J{=}0.04,h_{1}^{opt}/J{=}0.45, h_{2}^{opt}/J{=}{-}0.1\}$ and $\{h_{0}^{opt}/J{=}26,h_{1}^{opt}/J{=}0.2, h_{2}^{opt}/J{=}{-}0.5\}$, respectively, for $\bm{S1}$ and $\bm{S2}$.}\label{fig:Twoway1}
\end{figure}
\begin{figure}[t!]
	\centering\offinterlineskip
	\includegraphics[width=\linewidth]{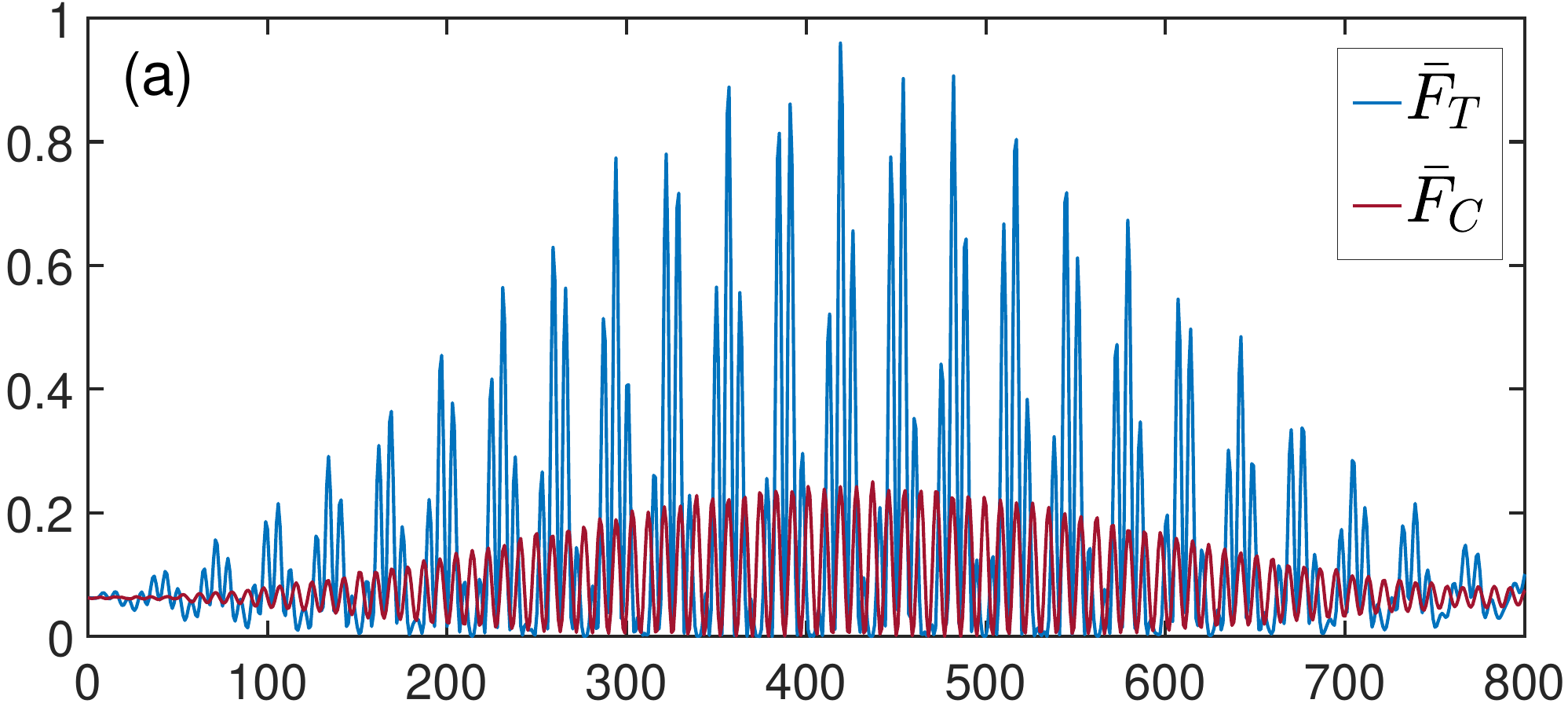}
	\includegraphics[width=\linewidth]{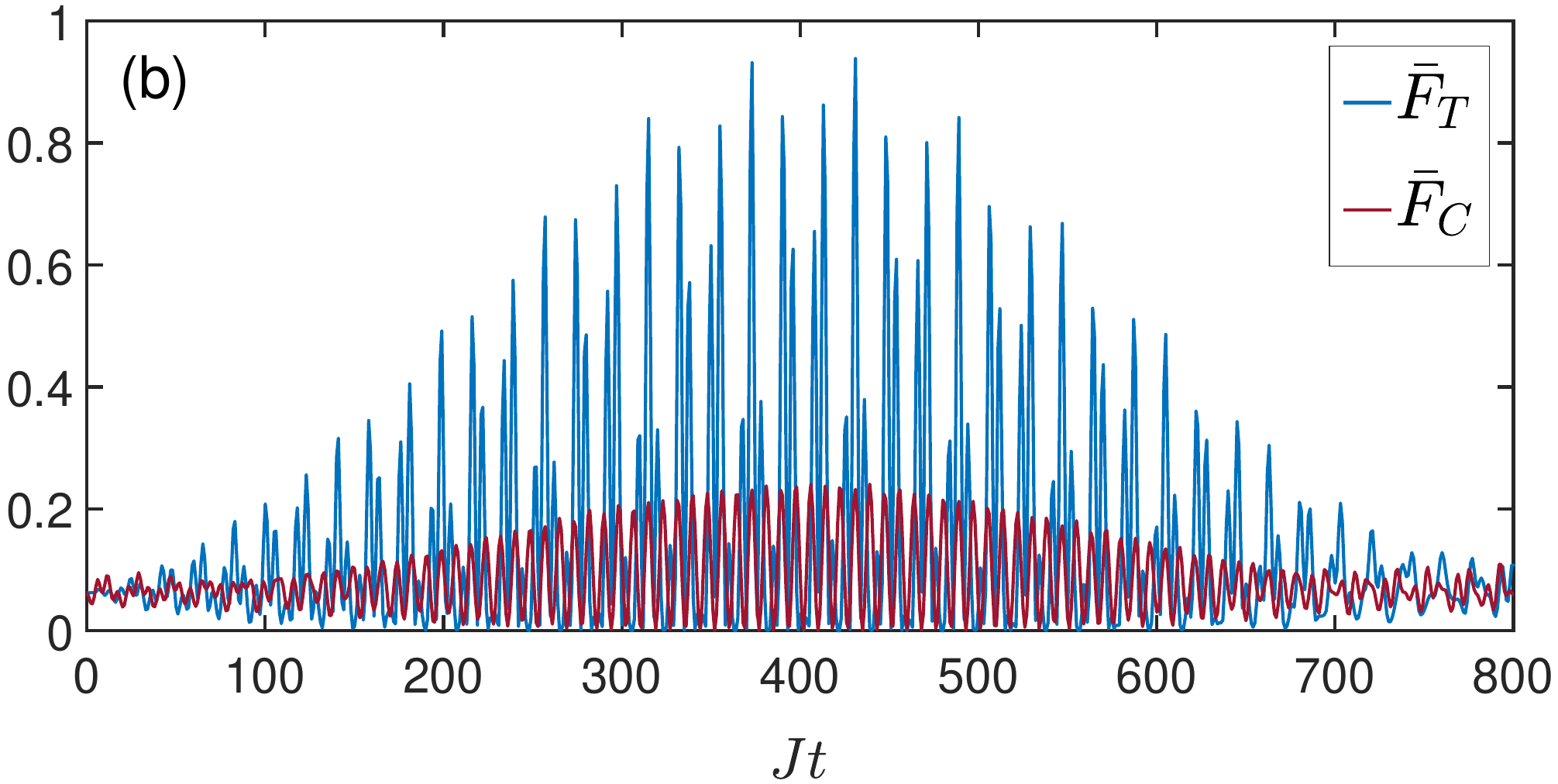}
	\caption{Two-way communication: The average transmission fidelity $\bar{F}_{T}(t)$ and the corresponding crosstalk $\bar{F}_{c}(t)$ for strategy $\bm{S1}$ (a) and $\bm{S2}$ (b) as a function of time in a chain of $N{=}10$. The Hamiltonian parameters are taken as $\{J_{0}^{opt}/J{=}0.04,h_{1}^{opt}/J{=}0.45, h_{2}^{opt}/J{=}{-}0.1\}$ and $\{h_{0}^{opt}/J{=}26,h_{1}^{opt}/J{=}0.2, h_{2}^{opt}/J{=}{-}0.5\}$ for $\bm{S1}$ and $\bm{S2}$, respectively.}\label{fig:Twoway2}
\end{figure}

 	
In Fig.~\ref{fig:Twoway1}(a) and (b) the maximum transmission fidelity $F_{T}^{max}(\tau)$ and corresponding $F_{C}(\tau)$ as functions of the polar angles $\theta_{1}$ and $\theta_{2}$ for our first strategy, $\bm{S1}$, in a chain of $N{=}10$ and optimal arriving time $\tau{\in} [1,500]/J$ are plotted. 
The same quantities for second strategy is plotted in Fig.~\ref{fig:Twoway1}(c) and (d). 
Here, the azimuthal angels $\phi_{1}$ and $\phi_{2}$ are considered as random numbers within $[0,2\pi]$.
Note that, in providing these plots, the Hamiltonian parameters are adjusted on their optimal values, resulting in maximal transmission fidelity and minimal crosstalk.
The most clear trend that can be identified from these plots is that, except for the senders' states on the southern hemisphere of the Bloch sphere, i.e., $\theta_{1},\theta_{2}{\in}[\pi/2,\pi]$, our protocol for other states results in high-fidelity transmission and low crosstalk.
The main reason for obtaining such results is our special choice of the channel's state in which all spin qubits are initialized at the north pole, $\vert 0 \rangle$, of the Bloch sphere.
This creates bias for the transmission of quantum state at the northern hemisphere of the Bloch sphere.  
Regarding the initial states of the receivers, for          senders' states near to the north pole of the Bloch sphere, i.e., $\theta_{1},\theta_{2}{<}\pi/4$, one obtains crosstalk more than $0.5$, the red triangles in the left corners of Fig.~\ref{fig:Twoway1}(b) and (d), due to the less distinguishability in  the states of the senders and receivers.
Obviously, our two strategies for implementing two-way quantum communication show mostly analogous behaviors for different types of initial states.
In Fig.~\ref{fig:Twoway2}(a) and (b), we plot the dynamics of average transmission fidelity  $\bar{F}_{T}$ and corresponding crosstalk $\bar{F}_{C}$ in a chain of length $N{=}10$, by adopting strategy $\bm{S1}$ and $\bm{S2}$, respectively. As the figures show, regardless of the adopted strategy, the average transmission fidelities  after some fluctuations reach to their highest values at a specific time $\tau{\in}[1,500]/J$ while the crosstalks remain negligible. 
  
\section{Conclusion}
In order to be universal quantum computers, digital quantum simulators require the capability of performing single-qubit gates and two-qubit entangling operations.
While the implementation of single-qubit gates relies on the capability of controlling individual particles and performing local unitary operations, the fulfillment of two-qubit entangling gates demands direct interaction between qubits which makes it more challenging, particularly, between distant qubits.
One of the attractive approaches to mediate the interaction between remote qubits and realize a two-qubit entangling gate is to employ
a spin chain data-bus and exploit the  non-equilibrium dynamics of the system. 
In most of this type of gate implementation, only one gate can be performed at each time. This effectively restricts the computational power of the quantum processors through limiting the depth of the circuits.             
To overcome this limitation, here, we have devised a protocol that is able to implement multiple two-qubit entangling gates in parallel on arbitrary pairs of distant qubits through a commonly shared spin chain data-bus.   
Remarkably, while our protocol implements entangling gate between several pairs of qubits simultaneously, it keeps the crosstalk negligible through making each pair of qubits off-resonant from the others by local tuning of the magnetic fields. We have put forward two different strategies to achieve these goals through optimizing different sets of Hamiltonian parameters. Each of these strategies might be more convenient for certain physical platforms. 
A remarkable feature of our proposal is that it is hardly affected by increasing the length of the data-bus and the number of users.
It should be emphasized that, our protocol is robust against  imperfections in fabrication and quantum noise effects. Therefore, while the optimization of the Hamiltonian parameters is important for functionality of the protocol, the obtainable quality is not very sensitive to the exact values of the parameters. To show the practicality of our protocol, we have proposed an experimental proposal based on  superconductor systems.
Finally, we show that the proposed protocol can be used for achieving  high-fidelity two-way quantum communication. \\
\section{Acknowledgement} 
AB acknowledges support from the National Key R\&D Program of China (Grant No.2018YFA0306703) and National Science Foundation of China (grants No.12050410253 and No.92065115).


\bibliographystyle{plain}

\end{document}